\documentclass[10pt, preprint]{emulateapj}

\usepackage{color}

\usepackage{style_wei_symbols}
\usepackage{style_wei}

\usepackage{natbib}
\begin{document}

\title{Chromospheric Jet and Growing ``Loop" Observed by {\it Hinode}: 
New Evidence of Fan-Spine Magnetic Topology Resulting From Flux Emergence}

	%


\author{Wei Liu\altaffilmark{1}\altaffilmark{2}, Thomas E.~Berger\altaffilmark{1}, Alan M.~Title\altaffilmark{1},
Theodore D.~Tarbell\altaffilmark{1}, and B.~C. Low\altaffilmark{3}}


\altaffiltext{1}{Lockheed Martin Solar and Astrophysics Laboratory, Department ADBS, 
Building 252, 3251 Hanover Street, Palo Alto, CA 94304}
\altaffiltext{2}{W.~W.~Hansen Experimental Physics Laboratory, Stanford University, Stanford, CA 94305}
\altaffiltext{3}{High Altitude Observatory, National Center for Atmospheric Research, P.O. Box 3000, Boulder, CO 80307}

\shorttitle{Chromospheric Jet Observed by {\it Hinode}}
\shortauthors{Liu et al.}

\journalinfo{Accepted by ApJ 2010 December}

\begin{abstract}	

We present observations of a chromospheric jet and growing ``loop" system that show new evidence of 
a fan-spine topology resulting from magnetic flux emergence. This event, occurring in an equatorial coronal hole
on 2007 February 9, was observed by the \hinode Solar Optical Telescope in unprecedented detail.
The predecessor of the jet is a bundle of fine material threads 
that extend above the chromosphere and appear to rotate about the bundle axis
at $\sim$$50 \kmps$ (period $\lesssim$$200 \s$). 
These rotations or transverse oscillations propagate upward at velocities up to $786 \kmps$.	
The bundle first slowly and then rapidly swings up,
with the transition occurring at the onset of an A4.9 flare. 
A loop expands simultaneously in these two phases (velocity: $16$--$135 \kmps$).	
Near the peak of the flare, the loop appears to rupture; simultaneous upward ejecta
and mass downflows faster than free-fall appear in one of the loop legs.
The material bundle then swings back in a whiplike manner and develops into a collimated jet, 
which is orientated along the inferred open field lines with transverse oscillations continuing at slower rates.
Some material falls back along smooth streamlines, showing no more 	
oscillations. At low altitudes, the streamlines bifurcate at 
presumably a magnetic null point and bypass an inferred dome, depicting an inverted-Y geometry. 
These streamlines closely match in space the late Ca loop 	
and X-ray flare loop.
These observations are consistent with the model that flux emergence in an open-field region
leads to magnetic reconnection, forming a jet and fan-spine topology.
We propose that the material bundle and collimated jet represent the {\it outer spine}
in quasi-static and eruptive stages, respectively,	
and the growing loop is a 2D projection of the 3D {\it fan surface}.
\end{abstract}


\section{Introduction}
\label{sect_intro}


Due to magnetic buoyancy \citep{ParkerE.mag-buoyancy.1955ApJ...121..491P}, 
magnetic flux ropes are expected to emerge from the convection zone 
into the corona through the photosphere and chromosphere. Such emerging flux regions 
\citep[EFRs;][]{Waldmeier.1st-flux-emerg.1937ZA.....14...91W, Ellison.1st-flux-emerg.1944MNRAS.104...22E}
give birth to sunspots and active regions \citep{WeartZirin.AR-birth.1969PASP...81..270W}.
When observed on the solar disk, an EFR is usually seen as a new bipole in magnetograms that
grows in size and magnetic flux \citep{ZwaanC.flux-emerg.1978SoPh...60..213Z}. 
The opposite polarities of the bipole separate from each other at typical velocities $\lesssim$$1\kmps$ 
\citep{HarveyK.MartinSF.EFR-separa-vel.1973SoPh...32..389H, ChouWang.EFR-separ-vel.1987SoPh..110...81C}.
Upflows of $\sim$$1\kmps$ at the photospheric level were observed
\citep{BrantsJ.EFR-blueshift-1km/s.1985SoPh...98..197B} and confirmed in recent MHD simulations
\citep{Archontis.emergeMHD.2004A&A...426.1047A, Martinez-Sykora_MHDemergenceII.2009ApJ...702..129M}.
Once in the low-$\beta$ corona, driven by its magnetic pressure, the emerging flux expands
at relatively larger velocities of 10--$20 \kmps$, as observed in rising \Ha arch filaments 
\citep{BruzekA.arch-filament.1967SoPh....2..451B, ChouZirin.arch-filament-rise.1988ApJ...333..420C} 
and extreme ultraviolet (EUV) loops \citep{Yashiro.Shibata.TRACE-EFR.2000ASPC..205..133Y} in EFRs.
Dense photospheric or chromospheric material dredged up by the emerging flux was 
found to consequently drain down the legs of arch filaments at 30--$50 \kmps$ 
\citep{BruzekA.arch-filam-drain.1969SoPh....8...29B, Roberts.arch-filam-drain.1970PhDT........19R}.
Such velocities of rise and drainage have been reproduced in MHD simulations 
\citep{Archontis.emergeMHD.2004A&A...426.1047A, FanY.rot-sunspot.2009ApJ...697.1529F}.

When a flux rope emerges into an open-field region (e.g., coronal hole), magnetic reconnection
between the emerging and ambient fields is expected to take place, 
producing a flare and material ejection \citep{HeyvaertsJ.flux-emerg-flare-model.1977ApJ...216..123H}.
Such ejections were observed as {\it surges} in \Ha
\citep{NewtonH.surge-discovery.1934MNRAS..94..472N, Roy.surge1973PhDT.........7R,
Kurokawa.Kawai.surge.flux-emerg1993ASPC...46..507K}
and as {\it jets} at other wavelengths, including white light 
\citep{WangYM.WL.EUV.jet1998ApJ...508..899W}, UV \citep{Brueckner.Bartoe.UVjet1983ApJ...272..329B}, 
EUV \citep{Alexander.Fletcher.jet.1999SoPh..190..167A}, 
and soft X-rays \citep{ShibataK.1st-SXT-jet.1992PASJ...44L.173S, StrongK.SXT-jet1992PASJ...44L.161S}.
A classification of standard and blowout jets was proposed \citep{Moore.jet-dichotomy.2010ApJ...720..757M}.
Torsional motions or helical features found in surges or jets
\citep{XuAA.surge-rotate.1984AcASn..25..119X, Kurokawa.untwist-filamt1987SoPh..108..251K,
ShimojoM.jet-stat.1996PASJ...48..123S, Patsourakos.EUVI-jet2008ApJ...680L..73P}
were interpreted as relaxation of twists from the emerging flux
\citep{Shibata.Uchida.helic-jet.1986SoPh..103..299S,
Canfield.surge-jet1996ApJ...464.1016C, Jibben.Canfield.twist-surge-stat2004ApJ...610.1129J}.
Numerical simulations have been extensively employed to explained various aspects of solar jets
\citep{Shibata.Uchida.helic-jetMHD.1985PASJ...37...31S, YokoyamShibata.jetModel1995Natur.375...42Y,
GalsgaardK.jet-emerg.2005ApJ...618L.153G, Nishizuka.giantCaHjet.2008ApJ...683L..83N,
Ding_MHD-emerg-jet.2010A&A...510A.111D}.

\begin{figure}[thbp]      
 \epsscale{0.7}	
 \plotone{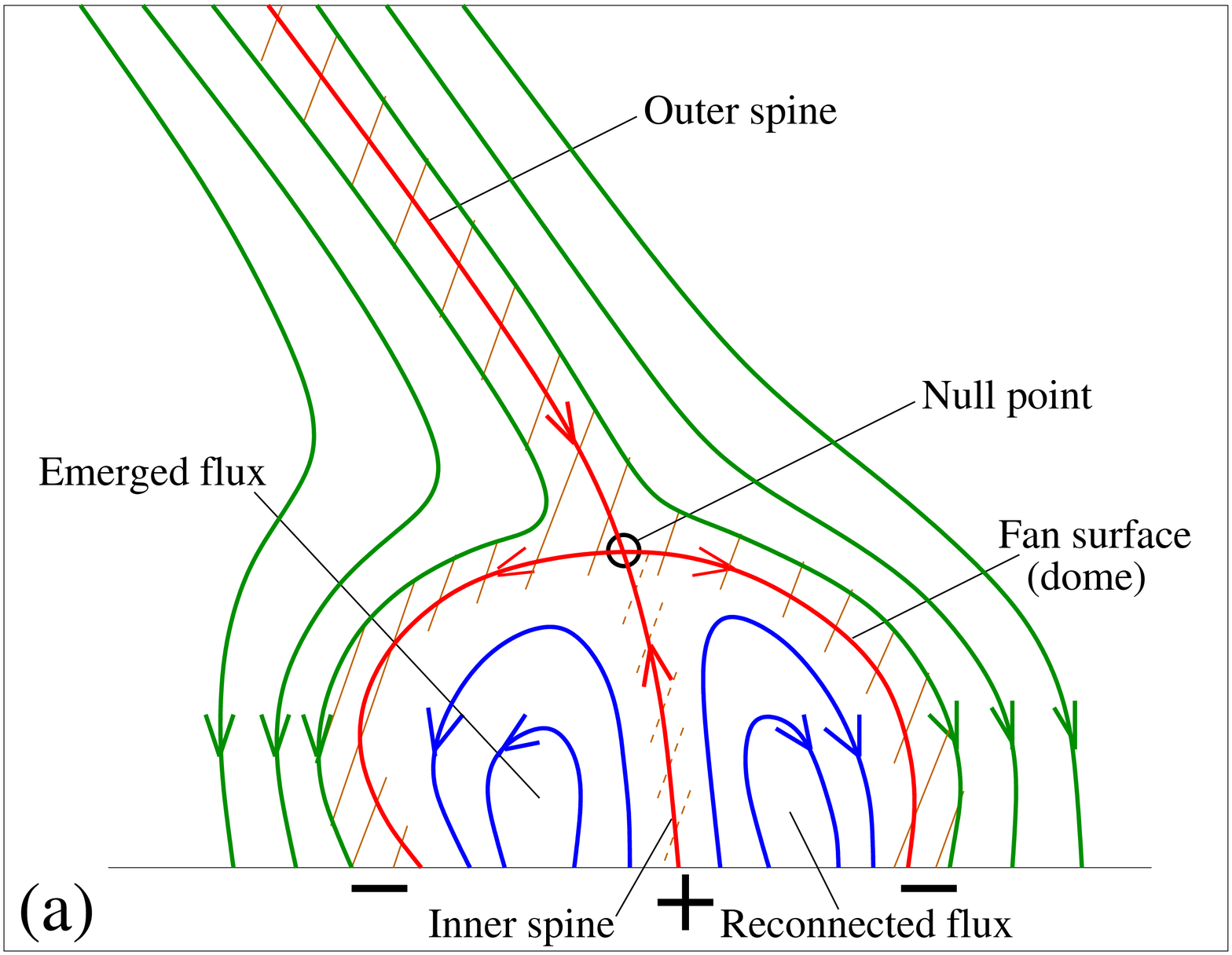}
 \plotone{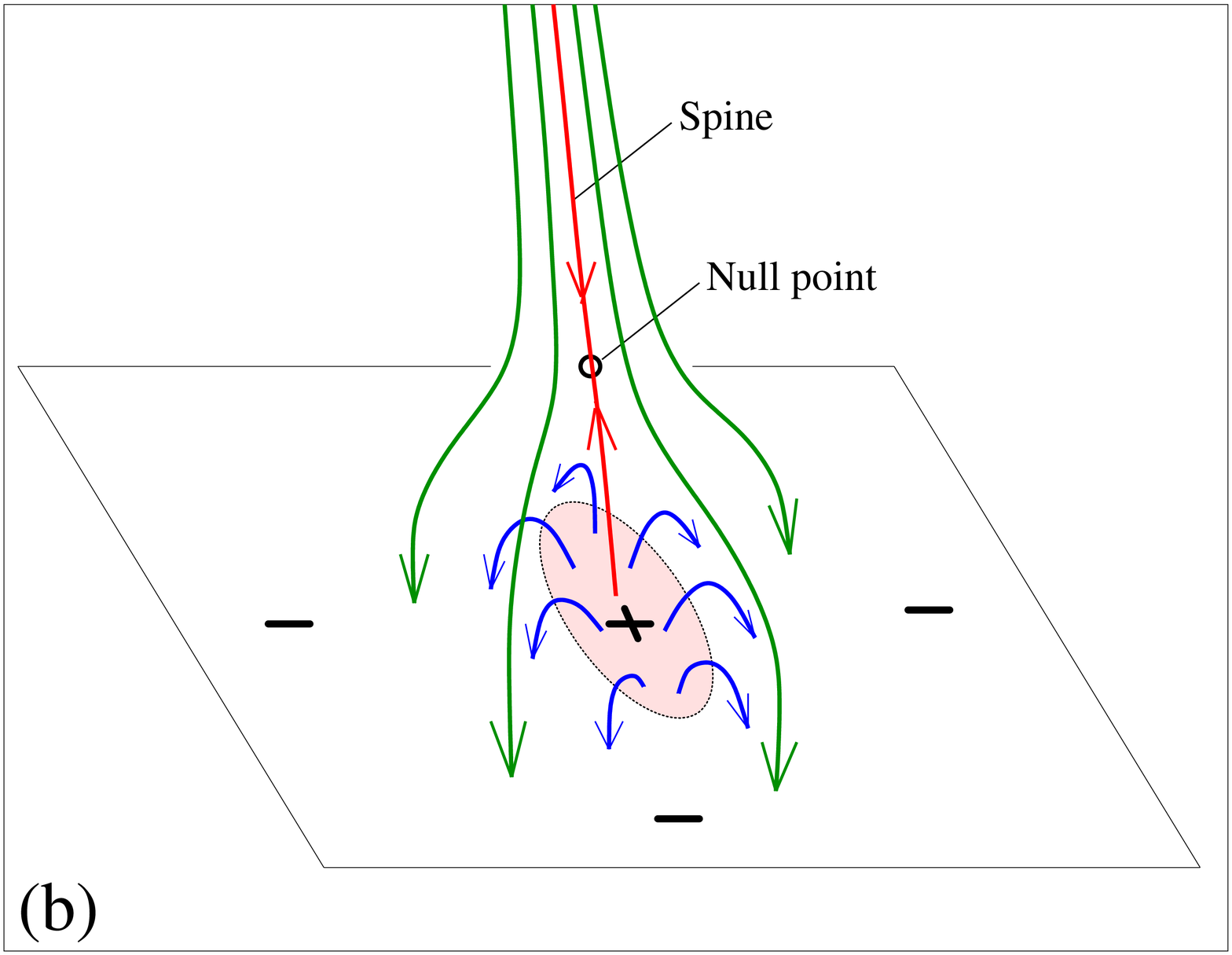}
 \caption[]{A fan-spine topology resulting from emergence of a bipole into a unipolar region: 
 (a) 2D vertical cut; (b) 3D bird's eye view.
 The hatched region in (a) represents postulated bright emission (see Section~\ref{sect_discuss}).
 } \label{fan-spine.eps}
 \end{figure}

The simplest end state of flux emergence into to a (locally) unipolar region is a {\it fan-spine} configuration
(\citealt{LauYT.Finn.3D-null.1990ApJ...350..672L, Torok.fan-spine.twist-emerg.2009ApJ...704..485T}; 
cf., multiple nulls connected by separators, \citealt{Maclean.null-emergence.2009SoPh..260..299M}).
As shown in Figure~\ref{fan-spine.eps}, 
consider a sufficiently small bipole emerging into a region
of a larger scale that has a net, say, negative, flux.
The emerged flux introduces two new patches of opposite polarities,
with the positive patch ending up as a minority-polarity isolated in a
negative polarity all around. Regardless of reconnection development,	
none of the field lines from this minority patch can leave this region, 
because the larger, surrounding region has a net opposite flux, 
and thus these field lines must fountain back to the nearby photosphere.
Immediately outside of this closed field is the open field,
and a {\it dome} or {\it separatrix fan surface} lies in between.
A magnetic {\it null point} is located on the top of this dome.  A special 
open field line, the {\it separatrix spine}, passes through this null point to 
continue on into the dome interior and be rooted at the base of the 
atmosphere.  The two parts of this spine, called the outer and inner, 
are identified in Figure~\ref{fan-spine.eps}.
A fan-spine topology bears significant implications for solar eruptions \citep{AntiochosS.breakout.1998ApJ...502L.181A}
and its signatures were found in anemone-like active regions
\citep{AsaiA.anemone-AR.2008ApJ...673.1188A},
Eiffel tower shaped X-ray jets \citep{ShimojoM.jet-stat.1996PASJ...48..123S},
saddle-like loop structures \citep{FilippovB.3D-null.1999SoPh..185..297F}, and
circular flare ribbons \citep{MassonS.oval-ribbon-spine-fan.2009ApJ...700..559M}.
Its presence in coronal jets or flares has been confirmed by magnetic field extrapolations 
\citep{Fletcher.3D-reconn.2001ApJ...554..451F, Moreno-Insertis.EISjet2008ApJ...673L.211M}
and widely reproduced in 2D or 3D MHD simulations 
\citep[e.g.,][]{YokoyamShibata.jetModel1996PASJ...48..353Y, PariatE.twist-jet-homologous.2010ApJ...714.1762P}.
%


The launch of the \hinode mission \citep{Kosugi.Hinode2007SoPh..243....3K}
has offered new opportunities to study
the relationship and underlying physics of flux emergence, jets, and fan-spine topology
in unprecedented detail \citep[e.g.;][]{LiHui.Hinode-flux-emerg.2007PASJ...59S.643L,
Shibata.CaHjet.2007Sci...318.1591S, Okamoto.emerg-helical.2008ApJ...673L.215O, 	
Morita_jet-SOT-Hida-Ca-spec2010PASJ...62..901M}.	
In an earlier Letter \citep[][hereafter Paper~I]{LiuW.CaJet1.2009ApJ...707L..37L},
we reported an intriguing chromospheric jet observed by \hinode on 2007 February 9 
and focused on the fine structure kinematics of the jet itself. 
In this paper, we present a multiwavelength study of the entire event in greater detail. 
In Section~\ref{sect_obs}, we provide context observations, infer the unipolar magnetic environment, 
and investigate the associated flare.
In Section~\ref{sect_sotanalys}, we pay special attention to the 
material bundle as the predecessor of the jet, the accompanying growing loop system, 
and the inverted-Y shaped geometry suggested by the streamlines of falling jet material.
In Section~\ref{sect_discuss}, we propose that flux emergence in the unipolar region
gives rise to the formation of a fan-spine topology, in which we identify the observed material bundle and jet
as the outer spine and the growing loop as a 2D projection of the 3D fan surface.
We conclude this paper in Section~\ref{sect_conclude} and provide supplementary information	
in the Appendices.	
 \begin{figure}[thb]      
 \epsscale{0.9}  
 \plotone{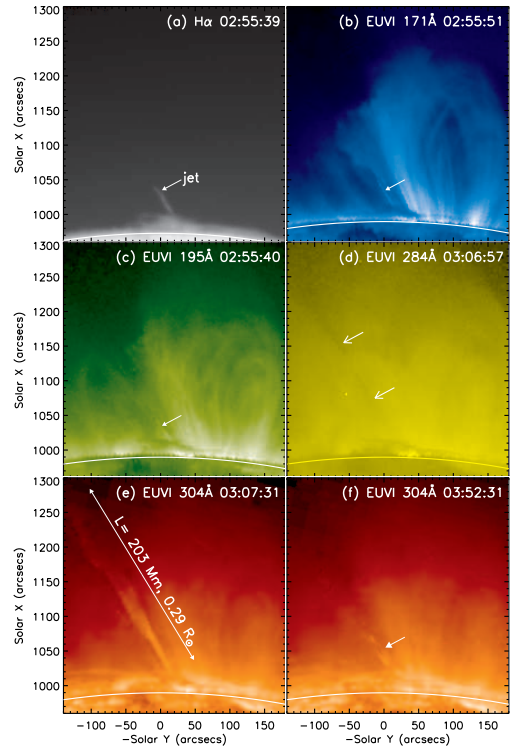}
 \caption[]{YNAO \Ha and various \stereo Ahead EUVI images at different stages of the event. 
 They are rotated by $90\degree$ counter-clockwise, such that the solar north is to the left.
 Note the dark absorption feature of the jet material in (d) and the bright falling material in (f).
 } \label{ha_stereo.eps}
 \end{figure}
\begin{table}[bthp]	
\scriptsize		
\caption{Event milestones.}
\tabcolsep 0.05in	
\begin{tabular}{ll}
\tableline \tableline
 02:14--02:30      &  earlier, brief surge-like activity  \\		
 02:32             &  bundle of material thread appears    \\		
 02:44             &  Ca loop and overarching SXR loop appear  \\
 02:49:02 ($t_1$)  &  onset of flare, and of fast rise of material bundle  \\
                   &  and Ca loop \\
 02:50:32 ($t_2$)  &  end of Ca loop lateral expansion \\
 02:51:12 ($t_3$)  &  material bundle's lower end turns from vertical \\
                   &  rise to horizontal drift; \\
                   &  Ca loop ``ruptures" (apex undetectable) \\
 02:51:44 ($t_4$)  &  elbow appears in material bundle; \\ 
                   &  northern Ca loop leg retreats downward \\
 02:52:24          &  material bundle apex starts to sweep northward \\
 02:52:40 ($t_5$)  &  orientation angle of material bundle axis reaches \\
                   &  maximum near flare peak; simultaneous upward  \\
                   &  ejecta and downflow in northern Ca loop leg \\
 02:55             &  Ca loop leg and overarching SXR loop invisible  \\
\tableline  \end{tabular}

\label{table_timeline} \end{table} 
%

\section{Multiwavelength Context Observations}
\label{sect_obs}

 \begin{figure*}[thbp]	
 \epsscale{0.95}	
 \plotone{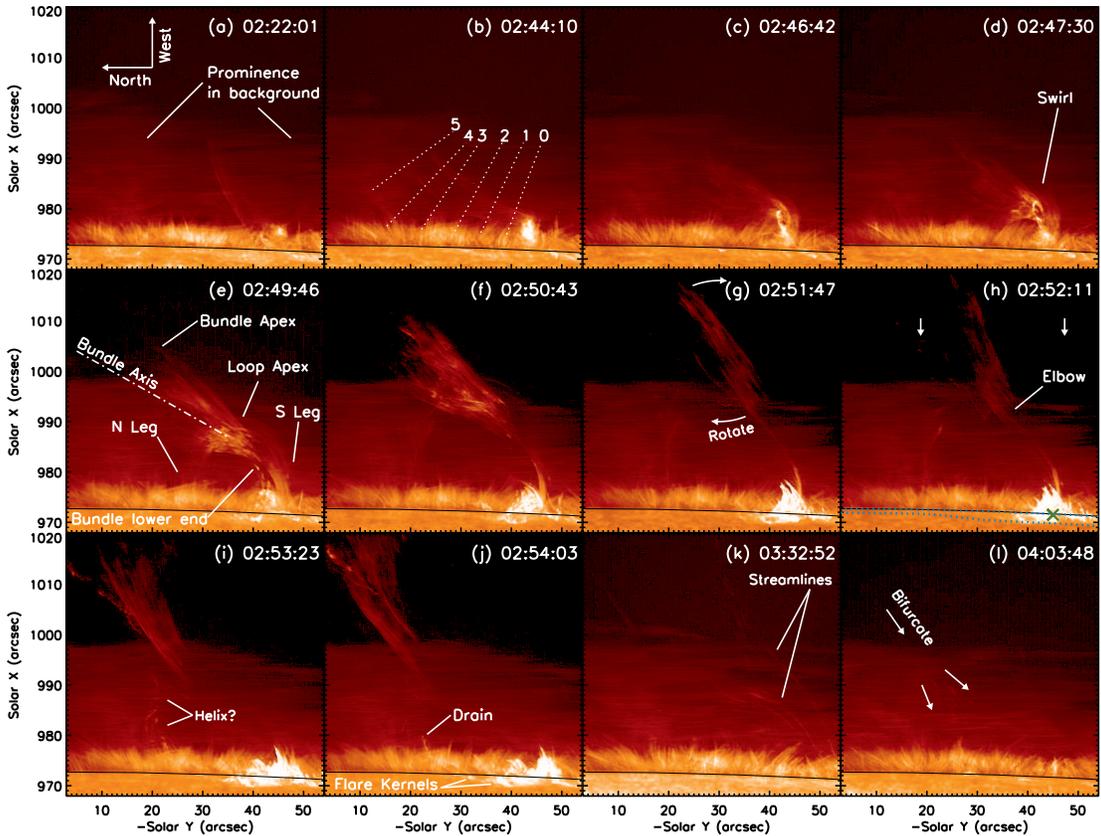}
 \caption[]{
 Sequence of \hinode \ion{Ca}{2} H images.	
 The numbered dotted lines in (b) represent $3\arcsec$ narrow cuts perpendicular to the local threads of the material bundle. 	
 The geometric elements of the event are marked in (e). In panel (h), the cross marks the central position of the
 flare and the dotted line delineates the boundary of the coronal hole 
 indicated by the PFSS model (see Figure~\ref{pfss_global.eps}); the two vertical arrows here point to
 the average, final solar-$y$ positions of the loop legs since $t_2=$~02:50:32~UT (see Figure~\ref{jet_vs_time.eps}(f)).
 Note the streamlines of the falling material in (k) and (l).		
 (See Animation~1 in the online journal and a higher resolution version at 
 \href{http://www.lmsal.com/~weiliu/public/hinode/2007-02-09_sot-jet}{http://www.lmsal.com/$\sim$weiliu/public/hinode/2007-02-09$\_$sot-jet (360 MB)}.)
 } \label{mosaic.eps}
 \end{figure*}
The event of interest occurred on the west limb near the equator at S03W89	
from 02:14 to 04:20~UT on 2007 February 9.
This region was active in producing jets, including one at 13:20~UT on the same day
\citep{Nishizuka.giantCaHjet.2008ApJ...683L..83N}.
The jet under study was observed by the \hinode Solar Optical Telescope \citep[SOT;][]{Tsuneta.SOT.2008SoPh..249..167T, Suematsu.OTA2008SoPh..249..197S}
in the \ion{Ca}{2}~H line, at a $0 \farcs 2$ spatial resolution and 8~s cadence.
It was also detected in \Ha by the Yunnan Astronomical Observatory (YNAO; see Figure~\ref{ha_stereo.eps}(a))
and a few other ground-based facilities, in EUV by \trace 
\citep{HandyB.trace.1999SoPh..187..229H}, \soho EIT, and \stereo EUV Imager 
\citep[EUVI;][Figures~\ref{ha_stereo.eps}(b)--(f)]{WuelserJ.STEREO-EUVI.2004SPIE.5171..111W}, 
and in soft X-rays (SXR) by the \hinode X-Ray Telescope \citep[XRT;][]{Golub.XRT.2007SoPh..243...63G}.
\hsi \citep{LinR2002SoPh..210....3L} and XRT also observed the accompanying \goes A4.9 flare in X-rays. 	
The procedure of \hinode SOT data reduction can be found in \href{http://adsabs.harvard.edu/abs/2009ApJ...707L..37L}{Paper~I} and details of coaligning
images from various instruments are provided in Appendix~\ref{Append_sect_coalign}.

\subsection{Overview of \ion{Ca}{2} H, EUV, and X-ray Observations}		
\label{subsect_multiObs}


%
 \begin{figure*}[thbp]      
 \epsscale{1.15}	
 \plotone{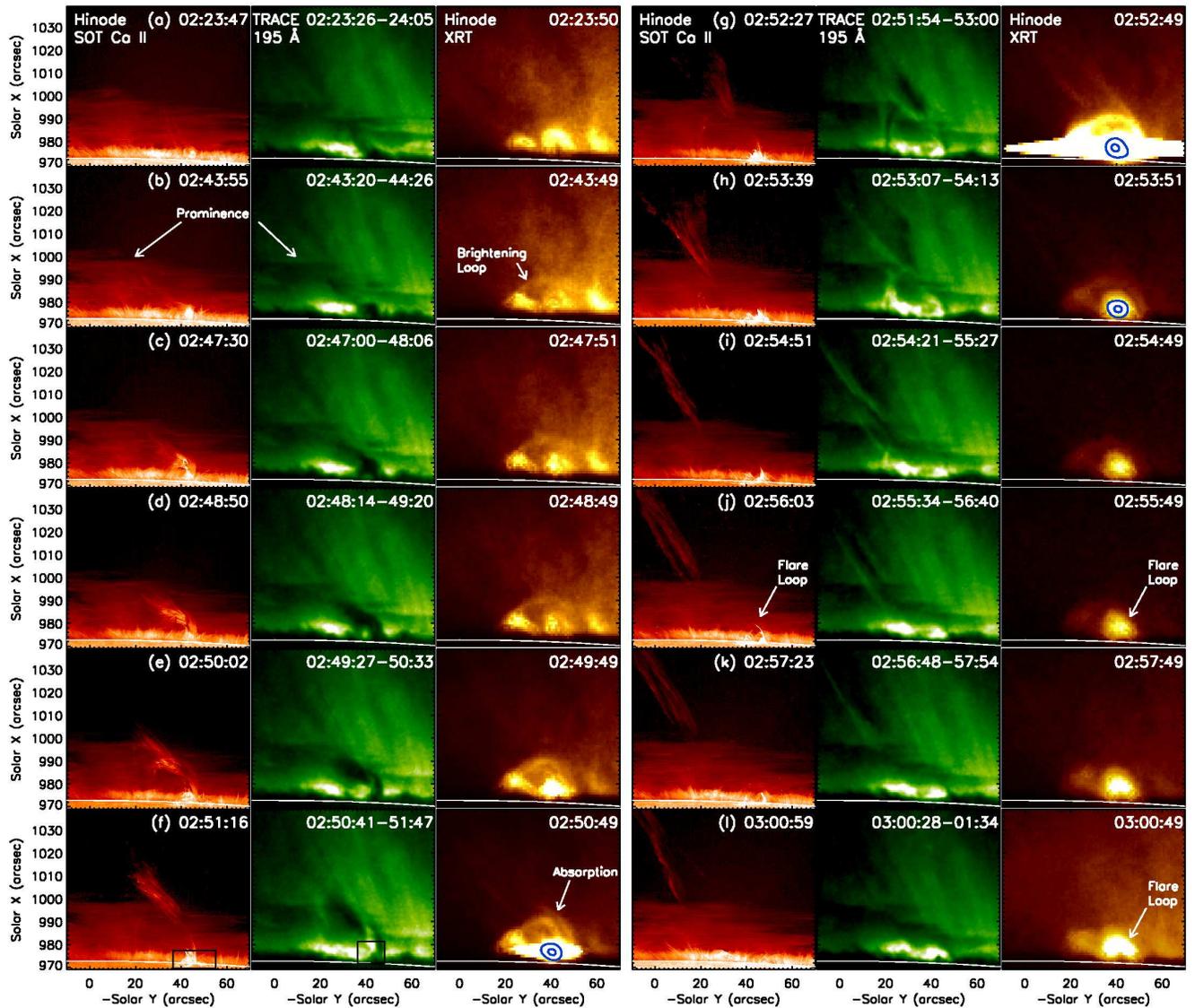}
 \caption[]{Nearly simultaneous images obtained by \hinode SOT in the \ion{Ca}{2}~H line (left),
 \trace at 195~\AA\ (middle), and \hinode XRT in SXR with the Al-poly or Al-poly+Ti-poly filter (right). 
 The \hinode images had exposure time on the order of 1~s,
 while \trace had long exposures as labeled. The contours at times (f)--(h) are \hsi 3--6~keV images
 with 100~s integration centered at the corresponding XRT time.	
 The boxes in panel (f) mark the regions for integrating source fluxes shown in Figure~\ref{flare_time_line.eps}(a).
 (See online Animation~2.)
 } \label{sot_trace_xrt.eps}
 \end{figure*}
Figures~\ref{mosaic.eps} and \ref{sot_trace_xrt.eps} show the evolution of the jet in the \ion{Ca}{2}~H line, EUV, 
and X-rays, and Table~\ref{table_timeline} summarizes the event time line.
Early in the event, from 02:14 to 02:30~UT, there is brief surge-like activity near the limb 
(Figure~\ref{mosaic.eps}(a))	
and tumbling motions are seen around its base at spicule heights.
At the same location, a bundle of material threads of typical width $\lesssim$$1\arcsec$ appears at 02:32~UT
(see online Animation~1, Figure~\ref{mosaic.eps}(b)). This bundle, as the direct predecessor of the later jet,
extends above the chromosphere at an oblique angle toward the north and gradually swings up clockwise, seemingly unfolding itself and
exhibiting oscillatory transverse motions across its axis.	
At the base of the bundle, a loop becomes visible at 02:44~UT and starts to expand both vertically and laterally (mainly toward the north).
From 02:46 to 02:48~UT, blobs of bright emission are seen to swirl up in a helical-like trajectory
from the chromosphere into the bundle (Figures~\ref{mosaic.eps}(c)--(d)).	
Most of these can also be seen in \trace 195~\AA\ images (green panels in Figure~\ref{sot_trace_xrt.eps})
at lower resolution ($1\arcsec$ vs.~$0 \farcs 2$ of SOT) as dark features, 
because the \ion{Ca}{2}~H line emission originates from partially ionized plasma at chromospheric
temperatures \citep[$\lesssim$$2\E{4}$~K;][]{Gouttebroze.prominence-CaII-lines.1997SoPh..172..125G}
that absorbs the background 195~\AA\ emission from hot plasma at coronal temperatures ($\sim$1.5~MK).
Some of these absorption features can be seen X-rays as well (e.g., Figure~\ref{sot_trace_xrt.eps}(f)).

 \begin{figure*}[thbp]      
 \epsscale{1.05}	
 \plotone{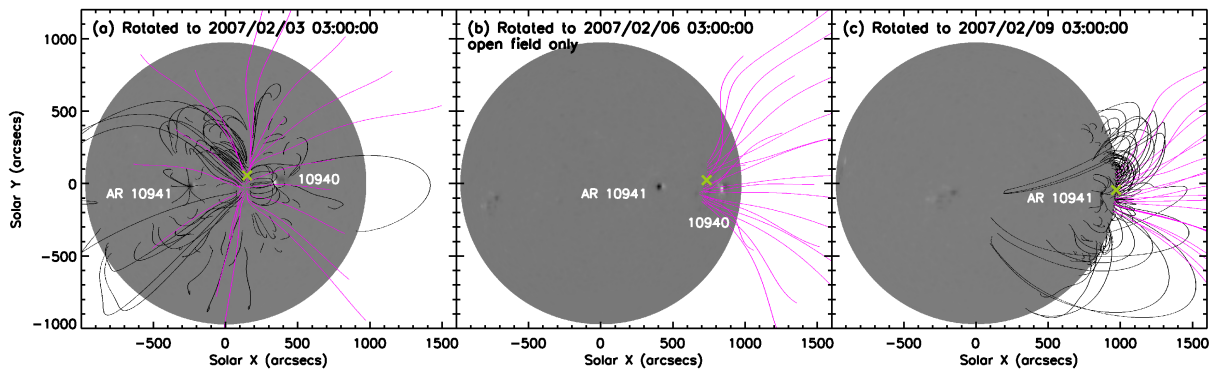}
 \caption[]{PFSS model \citep{Schrijver.DeRosa.PFSS.2003SoPh..212..165S} at 00:04~UT on 2007 February 9, 
 taken three hours before the jet. Each panel shows the global line-of-sight magnetogram and
 the field lines in the vicinity of the jet, as if they were seen from the Earth at the labeled time according to
 the Carrington rotation. Pink lines represent the open field of negative polarity, while dark lines are of the closed field.
 The cross sign is the same as that in Figure~\ref{mosaic.eps}(h), marking the central position of the \ion{Ca}{2}~H flare.
 } \label{pfss_global.eps}
 \end{figure*}
Around 02:48--02:49~UT, the swing and rise of the material bundle and the expansion of the loop start to accelerate.
At the same time ($t_1=$~02:49:02~UT), the A4.9 flare sets in near the lower	
southern leg of the loop. 
It appears as bright emission in the \ion{Ca}{2}~H line (Figure~\ref{mosaic.eps}), EUV, and X-rays (Figure~\ref{sot_trace_xrt.eps}).
At $t_2=$~02:50:32~UT, the loop legs stop their lateral expansion, and soon after ($t_3=$~02:51:12~UT), the loop appears to rupture and its apex
becomes undetectable (Figures~\ref{mosaic.eps}(f)--(h)). At $t_5=$~02:52:40~UT, the angle between the axis of the material bundle and the limb reaches its maximum.
The bundle then starts to sweep back toward the north in a whip-like manner and rapidly develops into a collimated jet,
which eventually extends to a length of 203~Mm or 0.29 $\Rsun$ (Figure~\ref{ha_stereo.eps}(e)).		
The loop continues to ``collapse" and material is seen to drain down the northern leg toward a flare kernel	
only $\sim$$15\arcsec$ away from the southern leg (Figures~\ref{mosaic.eps}(i)--(j)). This indicates that the lateral expansion of 
the loop occurs only in the corona, while the footpoints of the loop remain anchored on the surface.

The jet material is seen as emission in the cool ($\sim$$10^4$~K) \ion{Ca}{2}~H (Figure~\ref{mosaic.eps}) 
and \Ha lines	 
and at 304~\AA\ ($\sim$$10^5$~K) and 171~\AA\ ($\sim$1~MK), as absorption in the hotter ($\sim$2~MK) 284~\AA\ line 
(Figure~\ref{ha_stereo.eps}), as multiple strands of absorption and emission at intermediate 
($\sim$1.5~MK) 195~\AA\ (Figure~\ref{sot_trace_xrt.eps}),		
and as weak emission in XRT SXR ($\sim$1--30~MK; Figure~\ref{sot_trace_xrt.eps}(g)).
This indicates a wide range of temperatures, from $10^4$ to $10^6$~K, in the jet plasma.
The jet exhibits oscillatory transverse motions across its axis as seen by SOT, which we interpreted in \href{http://adsabs.harvard.edu/abs/2009ApJ...707L..37L}{Paper~I} as spins resulting from
unwinding twists. Further evidence of this interpretation is found at 195~\AA, where the emission and absorption
strands alternate and shadow each other in sequence, as expected for a spinning cylinder in a side view
(see online Animation~2).		

Later in the event, a flare loop appears in SXR and then EUV (Figure~\ref{sot_trace_xrt.eps}). Part of the jet material that
fails to escape the gravitational bound returns to the chromosphere along smooth streamlines in the original direction
of ascent (Figure~\ref{ha_stereo.eps}(f)). The oscillatory transverse motions seen earlier in the jet are no longer present in the falling material 
(see Figures~1 and 3 in \href{http://adsabs.harvard.edu/abs/2009ApJ...707L..37L}{Paper~I}).  At lower altitudes, most of the streamlines bypass a dome-shaped region and outline an inverted-Y geometry
(Figure~\ref{mosaic.eps}(k)).

\subsection{Magnetic Environment: Local Coronal Hole}
\label{subsect_mag}


This event occurred near the limb and thus there was no reliable direct magnetic field measurement available. 
We then resorted to the potential field source surface (PFSS) model 
\citep{ZhaoXP.Hoeksema.source-surface-model.1994SoPh..151...91Z, Schrijver.DeRosa.PFSS.2003SoPh..212..165S}
available in the {\it Solar SoftWare} package to obtain indirect information of the magnetic topology 
in the vicinity of the jet.  The modeled field is the extrapolation
of a photospheric magnetogram that combines \soho MDI observations within $60 \degree$ of the disk center and 
the evolved fields elsewhere according to the flux disperse model.

\begin{figure}[thbp]      
 \epsscale{1.6}	
 \plottwo{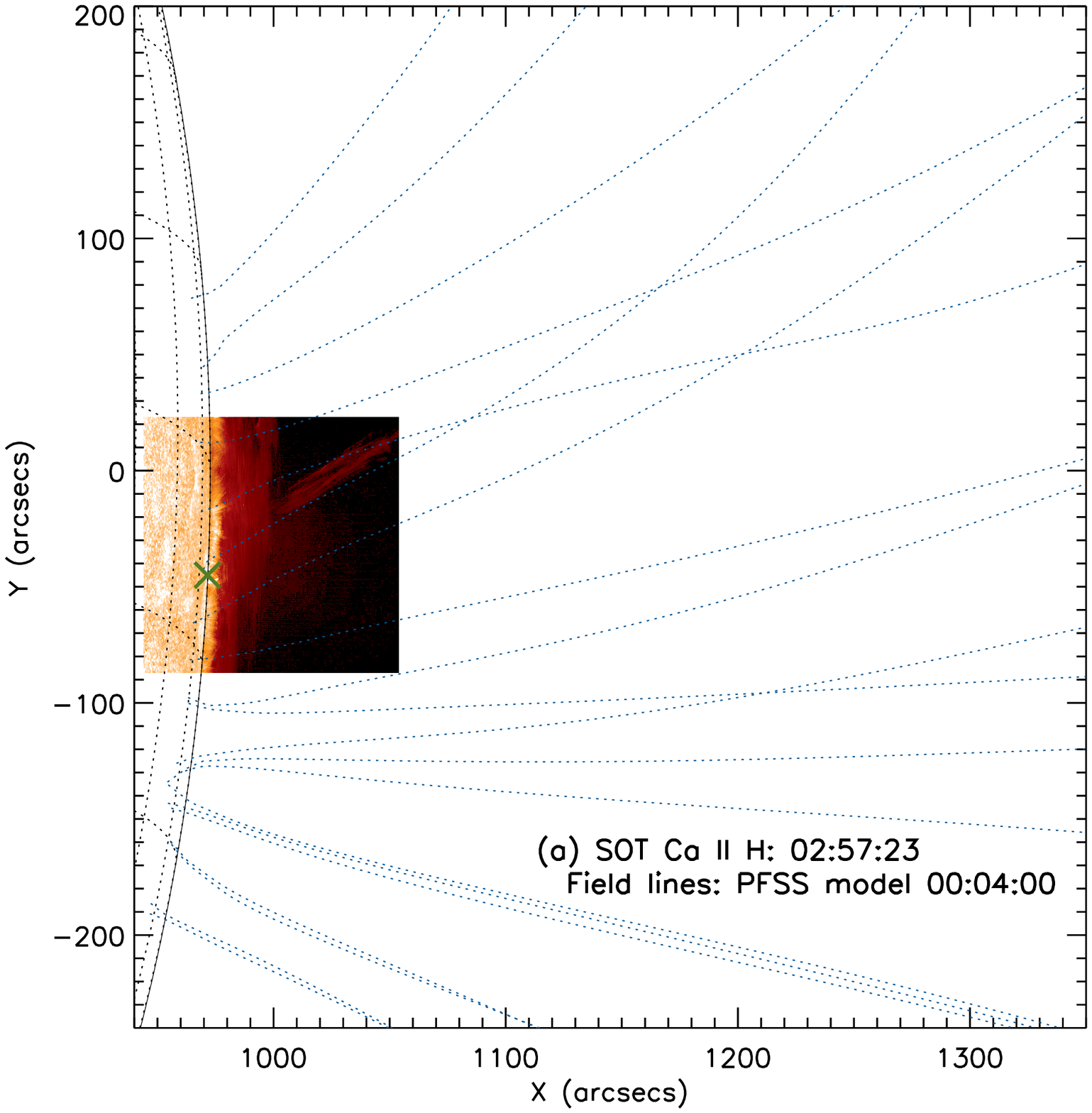}{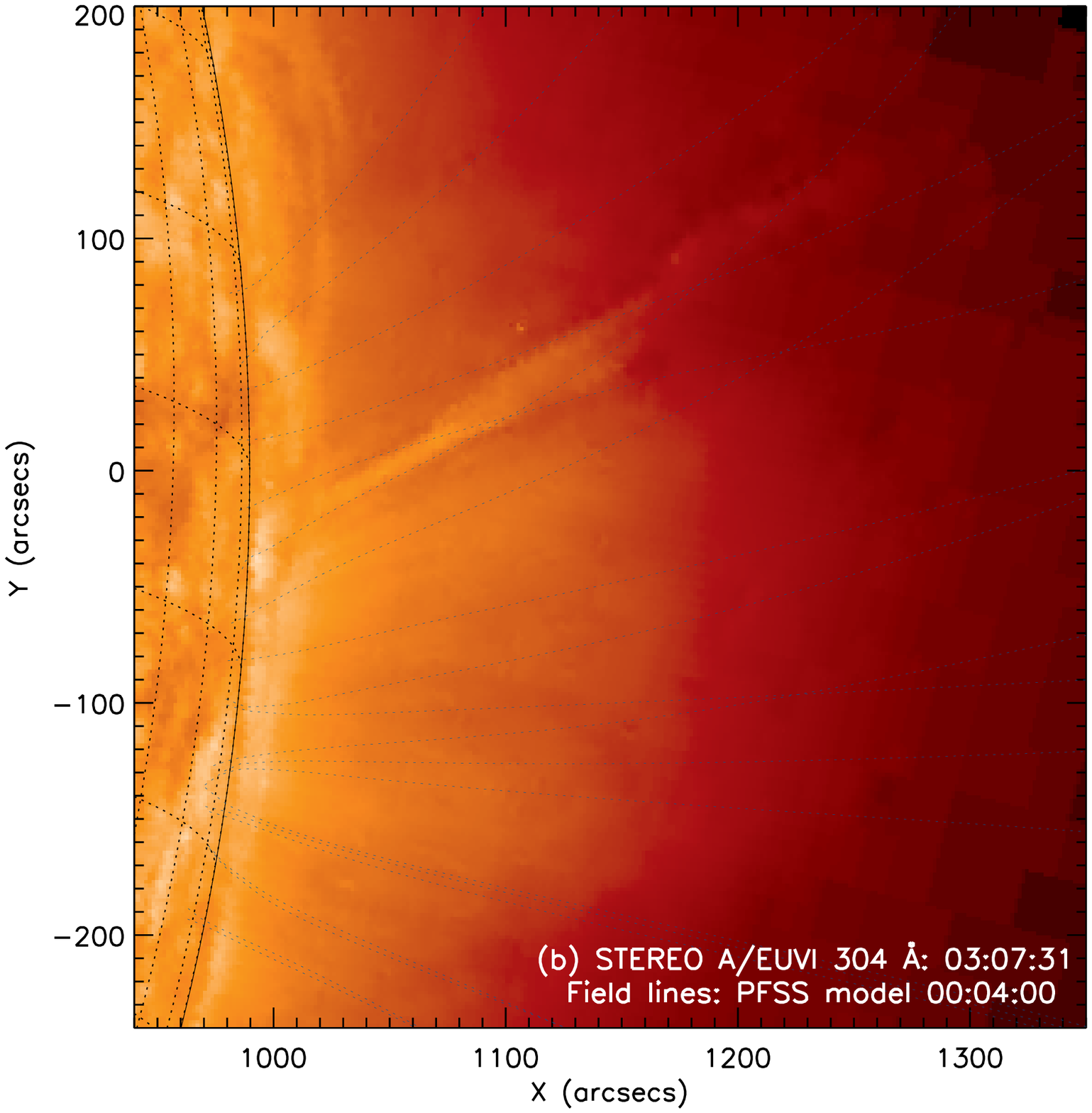}
 \caption[]{Open field lines of the PFSS model (as shown in Figure~\ref{pfss_global.eps}) overlaid on
 (a) the SOT image from Figure~\ref{sot_trace_xrt.eps}(k)	
 and (b) the \stereo Ahead 304 \AA\ image from Figure~\ref{ha_stereo.eps}(e).	
 The field lines were projected according to the positions of
 the corresponding spacecraft at the times of the images. The cross sign in (a) is the same as that in Figure~\ref{mosaic.eps}(h).
 } \label{sot_Bfield.eps}
 \end{figure}

Figure~\ref{pfss_global.eps} shows the PFSS model field at 00:04~UT near the time of the event, as seen from different view angles.
We noticed a north-south oriented narrow channel (range: $6\degree$ in longitude, $18\degree$ in latitude)
of open field or a {\it coronal hole} located between two NOAA active regions (ARs) 10940 and 10941.
 The models up to seven days prior to the event show that this coronal hole 
persistently existed, evolved slowly in size and shape,		
and matched the dark regions seen in \soho EIT 284~\AA\ images as expected.%
 \footnote{See the model predicted coronal hole boundaries at \href{http://www.lmsal.com/isolsearch}{http://www.lmsal.com/isolsearch}
 and their overlays on EIT images at \href{http://lmsal.com/forecast/modelEUV}{http://lmsal.com/forecast/modelEUV}.}
We then rotated the 00:04~UT model field to the time of the jet according to the Carrington rotation rate,
and found that the event was located inside the coronal hole. 
An example is shown in Figure~\ref{mosaic.eps}(h), where the coronal hole boundary (dotted line) encompasses 
the kernels	
of the flare. 
This is consistent with the finding of \citet{NittaN.jetHe3.2008ApJ...675L.125N} for an active region jet
and with the expectation that coronal jets tend to occur in open-field regions, as manifested by the ubiquitous polar jets
observed by \hinode \citep{Cirtain.XRTjet2007Sci...318.1580C, Shibata.CaHjet.2007Sci...318.1591S}.
Close comparison with the PFSS model further indicates that the north-west orientation of the jet is close to those of the
open field lines, again as expected from classical jet models. This can be clearly seen in both SOT \ion{Ca}{2}~H and \stereo
304~\AA\ images (Figure~\ref{sot_Bfield.eps}).
%

\subsection{Associated Microflare}
\label{subsect_flare}

\begin{figure}[thbp]      
 \epsscale{0.8}	
 \plotone{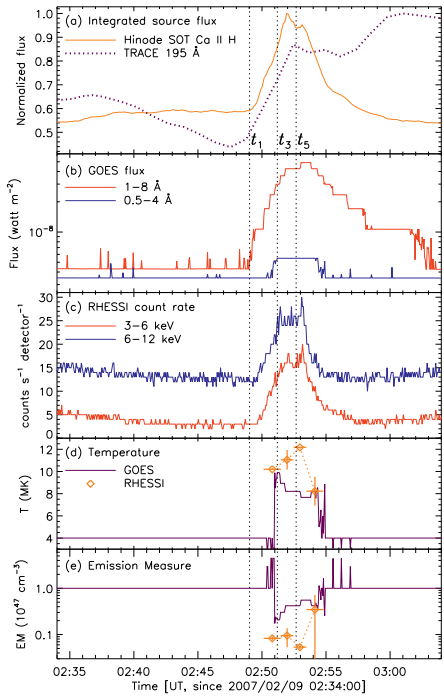}
 \caption[]{
 Time line of the accompanying flare.
 (a) Normalized \hinode \ion{Ca}{2}~H and \trace 195 \AA\ fluxes integrated over the source regions
 boxed in Figure~\ref{sot_trace_xrt.eps}(f).	
 (b) and (c) \goes and \hsi X-ray light curves.
 (d) and (e) Temperature and emission measure obtained from \goes and \hsiA.
 } \label{flare_time_line.eps}
 \end{figure}

The accompanying A4.9 flare occurred at 02:49~UT and peaked at 02:53~UT in the \goes low channel (1--8~\AA) flux.
As can be seen in Figures~\ref{flare_time_line.eps}(a)--(c), the light curves in the \ion{Ca}{2}~H line (SOT) and X-rays (\goes and \hsiA)
are very similar, showing a single hump, except for slightly different onset and peak times, presumably due to different 
instrument response to the varying plasma temperature. The \trace EUV light curve shows a different trend except during
the rise of the flare. The initial decrease from 02:37 to 02:47~UT is due to the unfolding material bundle and
growing loop at chromospheric temperatures, which lead to increased absorption of the background EUV emission.
The second flux increase (02:56--03:01~UT) is likely the result of the hot X-ray emitting plasma cooling to 195~\AA\ passband
temperatures and/or continuous evaporation of chromospheric plasma due to thermal conduction 
\citep{ZarroLemen.ConductionDriven.1988ApJ...329..456Z, Battaglia.conduct-evapor_2009A&A, LiuW.HDparticle-I.2009ApJ...702.1553L}.

The flare emission appears near the lower southern leg of the loop and cospatial at various wavelengths, from \CaIIH to EUV and X-rays
(Figures~\ref{mosaic.eps} and \ref{sot_trace_xrt.eps}). It is seen by \hsi up to the 6--12~keV band. 
Fits to \hsi spectra at the flare peak indicate that the emission is primarily thermal bremsstrahlung
from a plasma of temperature $T=12.2 \pm 0.6$~MK and emission measure ${\rm EM}=(5.3 \pm 0.5) \E{45} \pcmc$ 
(see Appendix~\ref{Append_sect_hsi-spec}). As shown in Figures~\ref{flare_time_line.eps}(d) and (e),
$T$ and ${\rm EM}$ have simple temporal profiles. Their differences between \goes and \hsi likely
result from the well-known fact that \goes is more sensitive to cooler plasmas.
Note that, because of their sharp occultation at the chromospheric limb,
the three bright loops seen earlier by XRT (Figure~\ref{sot_trace_xrt.eps}(a)) 
are most likely located in AR~10940 behind the limb, together with those large-scale EUV loops 
seen in Figure~\ref{ha_stereo.eps}.	
The flare loop happens to be in 
front of the background loop in the middle ($y \sim -40\arcsec$) and is located on the visible side 
of the disk, since its footpoints are inside the limb (Figure~\ref{sot_trace_xrt.eps}(l))
and the flare kernels	
are clearly observed (Figure~\ref{mosaic.eps}(j)).

\section{\hinode SOT Data Analysis}		
\label{sect_sotanalys}

In this section, we make geometric measurements of various features projected onto the sky plane
observed by \hinode SOT in the \ion{Ca}{2}~H line. Bear in mind that projection effects may limit
our capability of uncovering the true 3D picture. 
Here we assume that the displacements of Ca emission features represent motions of material rather 
than sequential excitations of emission due to temperature or density variations
(see Appendix~\ref{Append_sect_Ca-displace}).

\subsection{Material Bundle: Jet Predecessor and Early Development}
\label{subsect_bundle}

\begin{figure}[thbp]      
 \epsscale{1.0}	
 \plotone{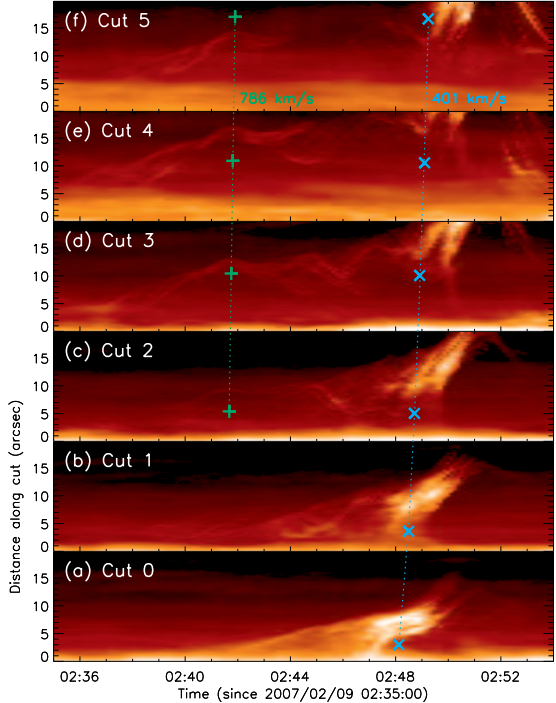}
 \caption[]{Space-time plot of cuts perpendicular to the local threads of
 the material bundle as shown in Figure~\ref{mosaic.eps}(b).
 The plus signs indicate the times of the crest of a 
 sinusoidal oscillation, and the crosses mark the onsets of the rapid
 rise of the bundle, both indicating a delay at higher cuts. 
 The vertical positions of the symbols are scaled with the separations of the
 corresponding cuts and are fitted with the dotted lines (see text) labeled by their final velocities.
 (See the original images overlaid with the cuts in online Animation~3.)
 } \label{tslice_perp.eps}
 \end{figure}
\begin{figure}[thbp]      
 \epsscale{1.0}	
 \plotone{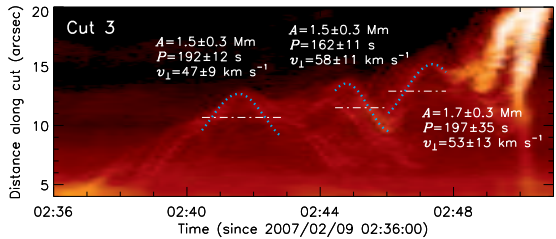}
 \caption[]{Same as Figure~\ref{tslice_perp.eps}(d) but with an enlarged view
 and slightly different color scale to emphasize oscillatory tracks.
 Overlaid are sinusoidal fits (thick dotted line) to selected tracks, and listed nearby
 are the fitted parameters: amplitude $A$, period $P$, and velocity $v_\perp$
 at the equilibrium position marked by the dot-dashed line.
 } \label{perp_fit.eps}
 \end{figure}

As briefly mentioned in Section~\ref{subsect_multiObs}, the predecessor of the jet is a bundle of material
threads that exhibits transverse motions across its axis. 
To quantify these motions, we have placed six $3\arcsec$ wide cuts along the bundle, oriented roughly perpendicular to 
the local threads as shown in Figure~\ref{mosaic.eps}(b). 
Space-time diagrams (Figure~\ref{tslice_perp.eps}) of these cuts clearly show oscillatory
transverse motions between 02:36 and 02:48~UT, particularly in the upper portion of the bundle 
(Cuts 2--5). We selected Cut~3 as an example and
fitted those unambiguously identified piecewise tracks with a sine function of time $t$:
 \beq 
 s_\perp(t) = s_{\perp 0} + A \sin \left[ {2 \pi (t-t_0) \over P } \right]  \,,
 \label{sin_eq} \eeq
where $s_\perp$ is the distance along the cut ($\perp$ to threads),
$A$ the amplitude, and $P$ the period. The resulting fits are shown in Figure~\ref{perp_fit.eps}.
Compared with those in the late stage of the jet as shown in Figure~4 of \href{http://adsabs.harvard.edu/abs/2009ApJ...707L..37L}{Paper~I},
the oscillation velocities ($v_\perp \sim 50 \kmps$) at the equilibrium position here are about	
1.5--2 times larger, while the amplitudes and periods are smaller by similar fractions.%
 \footnote{The amplitudes and periods might have been underestimated here, because each fit was done
 for less than a full cycle due to ambiguities and the changing brightness of
 material threads. However, the instantaneous oscillation velocities are not affected.}
%
%
To track the propagation of these oscillations, we identified the crests of an oscillation in Cuts 2--5
as marked by the plus signs on the left in Figure~\ref{tslice_perp.eps}, which exhibit a delay
from one cut to the next. We then fitted the separations
of the cuts vs.~the occurrence times of the crests with a linear function. (Here the cut separations 
are defined as the distance from the center of the cut to the previous neighboring cut.)
This yields a speed of $v_{\rm ph} =786 \pm 30 \kmps$ for the upward propagation, which is 
formally identified as a phase speed here.
This speed is similar to its later counterpart at the
leading edge of the jet (see Figure~3 of \href{http://adsabs.harvard.edu/abs/2009ApJ...707L..37L}{Paper~I}).  If these oscillations are MHD	
waves, the wavelength would be $\lambda = v_{\rm ph} P = v_{\rm ph}(192 \pm 12 \s) = 151 \pm 11 \Mm$.
We also located the onset (marked by crosses) of the rapid rise of the bundle in each cut.
A parabolic fit indicates an acceleration of $a= 4.6 \pm 0.5 \km \pss$ and a final speed of
$v_{\rm ph} =401 \pm 39 \kmps$ at the highest Cut~5.

Next, we quantify the rise of the material bundle by tracking its apex, lower end, 
and axis as marked in Figure~\ref{mosaic.eps}(e). 
The apex and lower end are identified as the highest and lowest point in altitude
of the ensemble of the bundle threads, respectively.
The axis is recognized as the bisector of the bundle's angular extent and is characterized by its 
orientation angle $\theta_{\rm bdl}$ from the limb. Figures~\ref{jet_vs_time.eps} shows the history 
of these geometric parameters, which, together with those shown in Figure~\ref{NLeg.eps},
are fitted with a piece-wise linear and/or parabolic function of time, depending on their temporal evolution,
to infer their velocities and/or accelerations.

  We find a {\bf slow-to-fast two-phase} evolution clearly divided by the onset of the accompanying flare at $t_1=$~02:49:02~UT.
 (1) Prior to $t_1$, the apex height $h_{\rm bdl}$, axis angle $\theta_{\rm bdl}$, and displacements of the bundle's lower end
from its initial position	
($\Delta x_{\rm bdl}$, $\Delta y_{\rm bdl}$) show {\it slow} variations (Figures~\ref{jet_vs_time.eps}(b)--(d)).
The clockwise upward swing of the axis is evident in the increase of $\theta_{\rm bdl}$ at a
moderate velocity of $\dot{\theta}_{\rm bdl}= (4.9 \pm 0.2) \E{-2} \, {\rm deg} \ps$.
 (2) At the flare onset, all of these quantities but $\Delta y_{\rm bdl}$ start to increase {\it rapidly}. 
$\dot{\theta}_{\rm bdl}$ experiences an acceleration 			
and reaches $(3.1 \pm 0.2) \E{-1} \, {\rm deg} \ps$
at $t_5=$~02:52:40~UT when its growth comes to a stop followed by a reverse. 
Overall, $\theta_{\rm bdl}$ has gained more than $50 \degree$ in 10~minutes.
The bundle's lower end first moves mainly upward ($\Delta x_{\rm bdl}$) with an acceleration of 
$0.66 \pm 0.09 \km \pss$ (final velocity: $110 \pm 6 \kmps$). At $t_3=$~02:51:12~UT,
it {\bf turns} to a primarily horizontal direction ($\Delta y_{\rm bdl}$, {\it northward}) with a comparable acceleration
and slight decrease in height by $3 \arcsec$ (Figures~\ref{jet_vs_time.eps}(d) and \ref{streamline.eps}(c)). 
At the same time, the top portion of the bundle continues swinging up and {\it southward}.%
  \footnote{The bundle's apex height does not appear to increase monotonically,
  because of the jumps caused by changes in brightness and thus switches of the feature to be tracked
  (Figure~\ref{jet_vs_time.eps}(b)).}
This makes the bundle rotate clockwise, forming an {\bf elbow} or bend toward the north at $t_4=$~02:51:44~UT 
(Figures~\ref{mosaic.eps}(g) and (h)). Soon after this (02:52:24~UT), 
the {\it entire} material bundle quickly sweeps toward the {\it north} like a whip, developing into a collimated jet,
and individual threads stretch upward as fast as $176 \pm 11 \kmps$ (Figure~\ref{jet_vs_time.eps}(b)).

	%
	%
	%
	%
	%
	%

\subsection{Accompanying Growing Loop}
\label{subsect_loop}

 \begin{figure}[thbp]      
 \epsscale{1.05}	
 \plotone{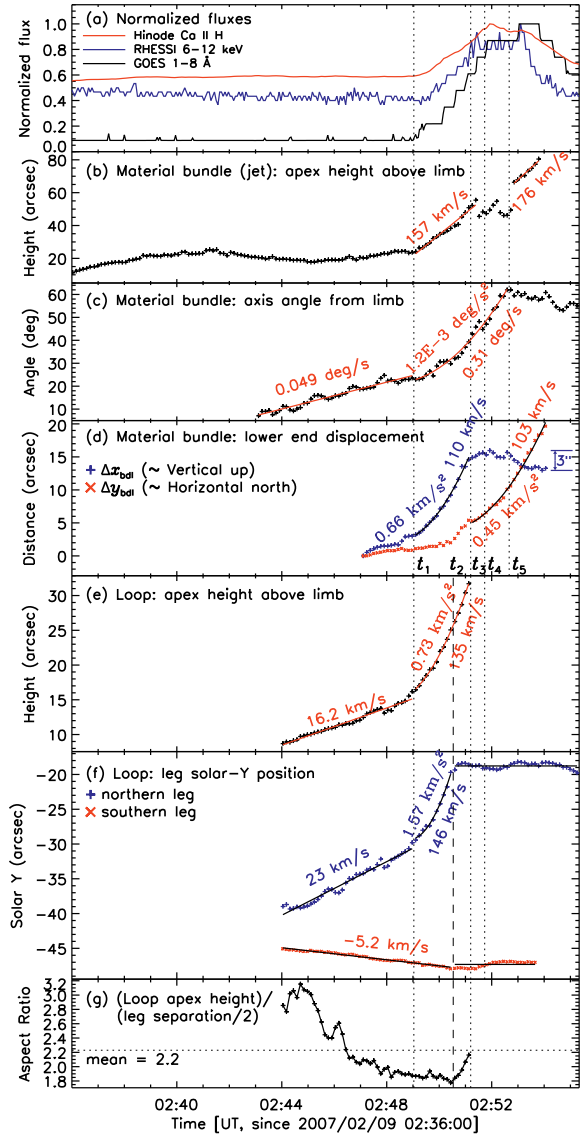}
 \caption[]{History of geometric quantities of the \ion{Ca}{2}~H material bundle (jet) and loop.
  (a) Light curves at various wavelengths repeated from Figure~\ref{flare_time_line.eps}.
  (b) Height $h_{\rm bdl}$ of the material bundle apex.
  (c) Angle $\theta_{\rm bdl}$ of the material bundle axis from the horizon.
  (d) Solar $x$ and $y$ displacements of the lower end of the material bundle as shown in
 Figure~\ref{streamline.eps}(c).
 Major milestones of the event are labeled as $t_1$, ..., $t_5$.
  (e) Height $h_{\rm loop}$ of the loop apex.
  (f) Solar-$y$ coordinates of the northern and southern legs of the loop,
 $y_{\rm NLeg}$ and $y_{\rm SLeg}$.
 The horizontal lines starting at $t_2=$~02:50:32~UT mark the means of the data after this time.
 Velocities and accelerations from linear and parabolic fits are labeled in (b)--(f).
  (g) Aspect ratio of the loop apex height $h_{\rm loop}$ to the half separation
 of the two legs in the north-south direction, ($y_{\rm NLeg} - y_{\rm SLeg})/2$.
 } \label{jet_vs_time.eps}
 \end{figure}
Another key component of this event is the growing \ion{Ca}{2}~H loop that accompanies the rise and eruption
of the material bundle. We track the evolution of its apex height $h_{\rm loop}$ 
and the lateral span of its two legs (Figures~\ref{mosaic.eps}(e)). 
The latter is derived from the solar-$y$ coordinates of the legs' outer edges 
since the $y$-axis here near the equator is almost parallel to the limb.

Similar to the material bundle, the loop apex also experiences two distinct phases divided at
the flare onset $t_1$ (Figure~\ref{jet_vs_time.eps}(e)):		
  (1) a {\bf gradual phase} from 02:44 to $t_1$ with a moderate ascent velocity of 
$16.2 \pm 0.4 \km \ps$ 			
and (2) an {\bf acceleration phase} from $t_1$ to $t_3=$~02:51:12~UT with an acceleration of $0.73 \pm 0.06 \km \pss$	
and a final velocity of $135 \pm 4 \km \ps$.		
The second phase ends when the loop appears to rupture and its apex is no longer visible.
This coincides with the turning point of the material bundle's lower end mentioned above 
(Figure~\ref{jet_vs_time.eps}(d)).
Similar phases of slow and fast rises have also been reported in other eruptive
events, including coronal jets \citep{Patsourakos.EUVI-jet2008ApJ...680L..73P} and prominence eruptions
\citep{SterlingMoore.slow-fast.2005ApJ...630.1148S, ChiforC.filamentErupt.2006A&A...458..965C,
LiuW.filmnt.2009ApJ...698..632L}.


%
 \begin{figure*}[tbhp]      
 \epsscale{0.85}	
 \plotone{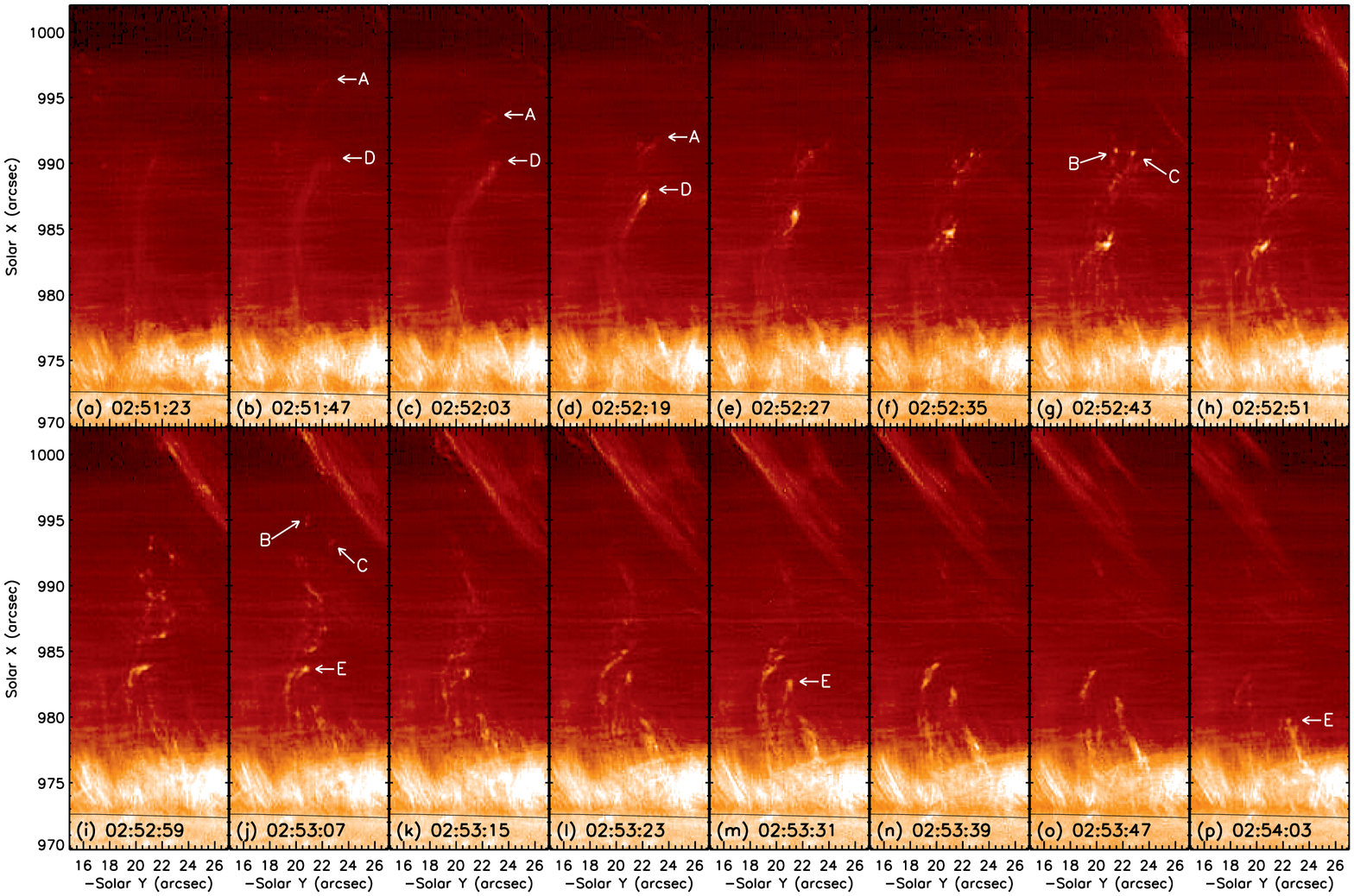}
 \caption[]{
 Enlarged view of the northern leg of the \ion{Ca}{2}~H loop at its late stage.
 Note the downward retreat of branches~A and D
 and the upward ejection of blobs~B and C.
 }\label{mosaic_NLeg.eps}
 \end{figure*}
 \begin{figure}[thbp]      
 \epsscale{1.0}	
 \plotone{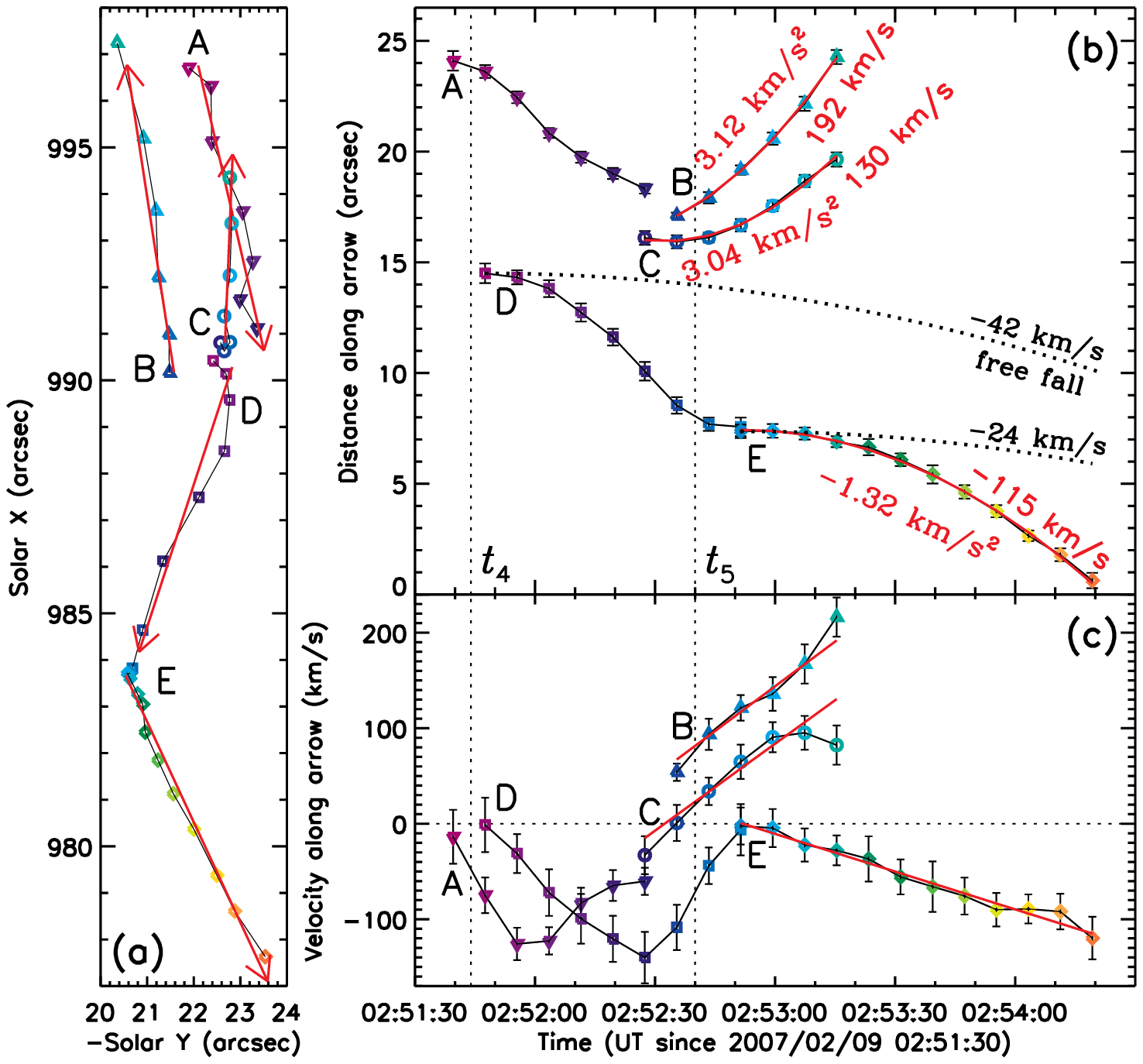}
 \caption[]{Kinematics of the emission blobs in the northern leg of the loop as marked in Figure~\ref{mosaic_NLeg.eps}.
   (a) Locations of blobs A--E with colors from magenta to red indicating time evolution.
 For each blob, a red arrow represents the main direction of motion, given by a linear fit to the data.
   (b) Distance along the main direction of motion for each individual blob, with positive (negative) for upward (downward).
 The curves are shifted vertically to make their end points approximately represent the relative heights
 of the corresponding blobs.
   (c) Time derivative of the distance in (b). 
 The red lines in (b) are parabolic fits to curves B, C, and E, labeled with the fitted accelerations and final velocities,
 and their counterparts in (c) are the corresponding velocities vs.~time.
 } \label{NLeg.eps}
 \end{figure}
The lateral position $y_{\rm NLeg}$ of the northern leg of the loop has a similar two-phase evolution with
1--2 times larger velocities and acceleration (Figure~\ref{jet_vs_time.eps}(f)). 
In contrast, the southern leg does not show such a distinction 
and has a uniform low velocity of $5.2 \pm 0.1 \km \ps$, only $1 \over 4$ of its northern counterpart
($23.3 \pm 0.7 \km \ps$) in the gradual phase. This {\bf asymmetry} means that the lateral expansion of the
loop is primarily on the northern leg moving toward the north.
Note that the lateral expansion velocity and acceleration averaged between the two legs
roughly equal their vertical counterparts.
The lateral expansion, however, ceases {\it earlier} than the vertical growth. 
This occurs in both legs at $t_2=$~02:50:32~UT~$<t_3$, after which they remain stationary
with small fluctuations (standard deviation: $0\farcs 4$).


%
 \begin{figure*}[bthp]      
 \epsscale{1.0}	
 \plotone{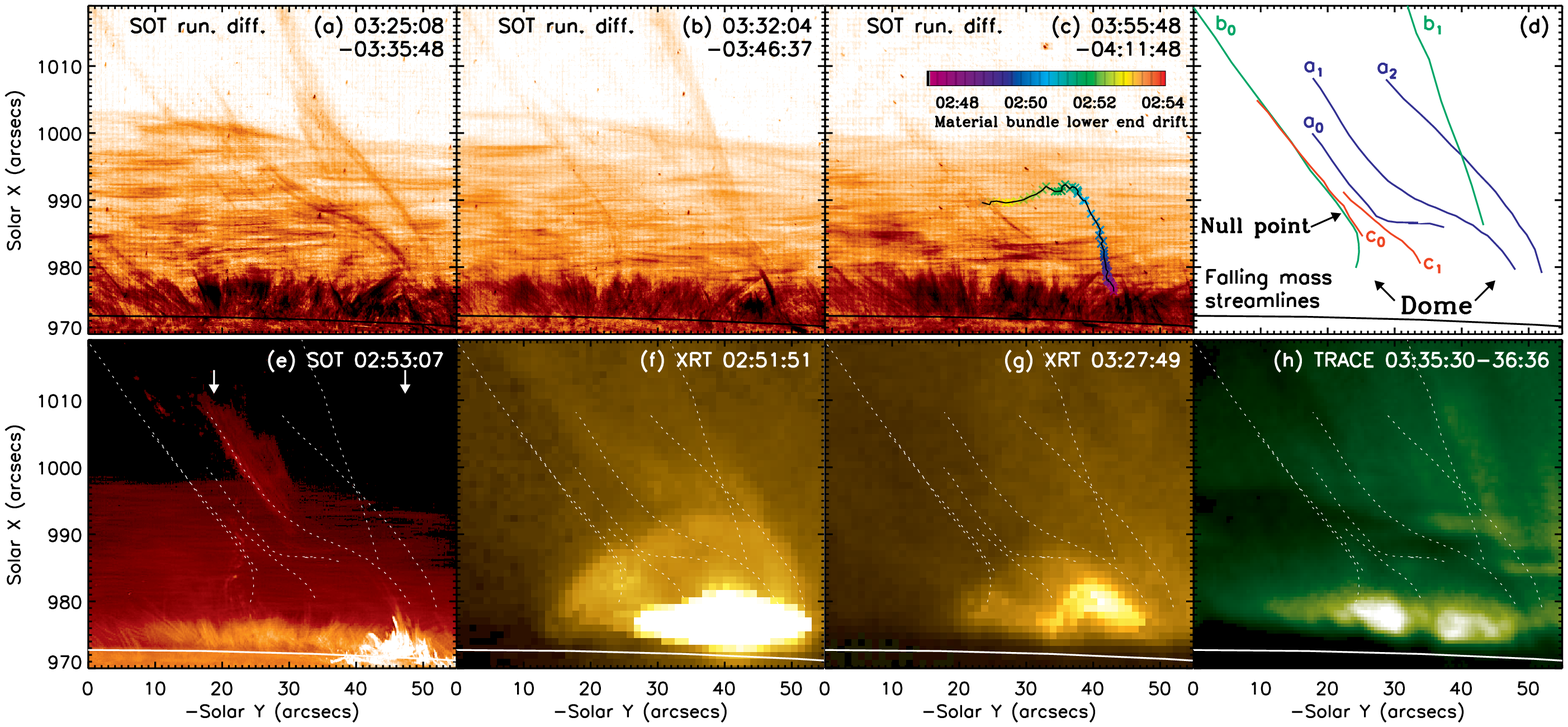}
 \caption[]{
  (a)--(c) Stacked running difference images showing streamlines of the falling material in three time intervals.
  (d) Streamlines obtained from panels (a)--(c) in different colors as labeled.
  (e)--(h) \hinode SOT \ion{Ca}{2}~H, XRT, and \trace 195 \AA\ images at various stages of the event, overlaid with
 the same streamlines shown in (d).
 The colored symbols in (c) represent the temporal migration of the lower end of the material bundle (see Section~\ref{subsect_bundle}),
 presumably near the null point.
 The two vertical arrows in (e) are the repetition from Figure~\ref{mosaic.eps}(h).
 } \label{streamline.eps}
 \end{figure*}
 \begin{figure*}[thbp]      
 \epsscale{0.27}	
 \plotone{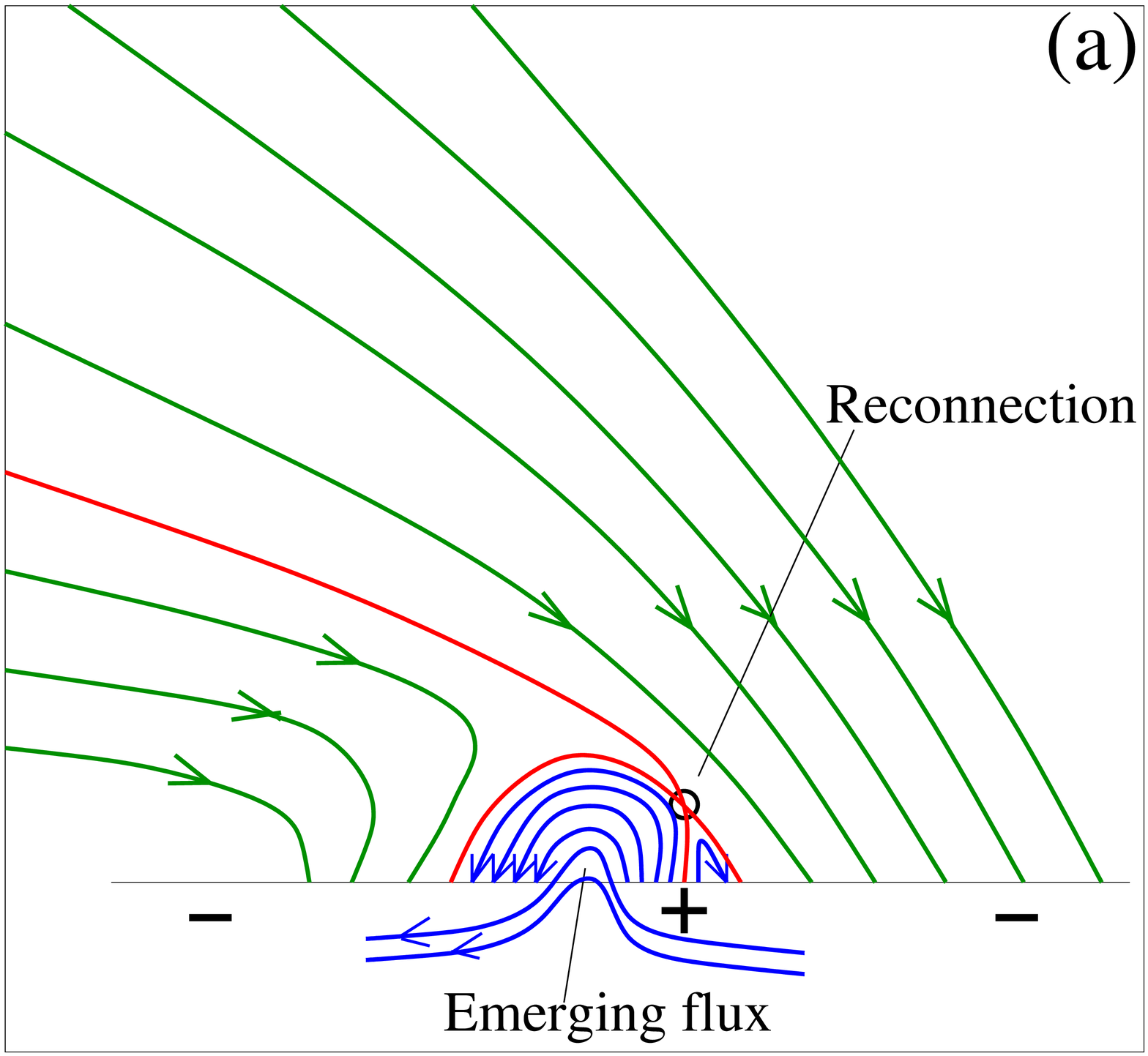}
 \plotone{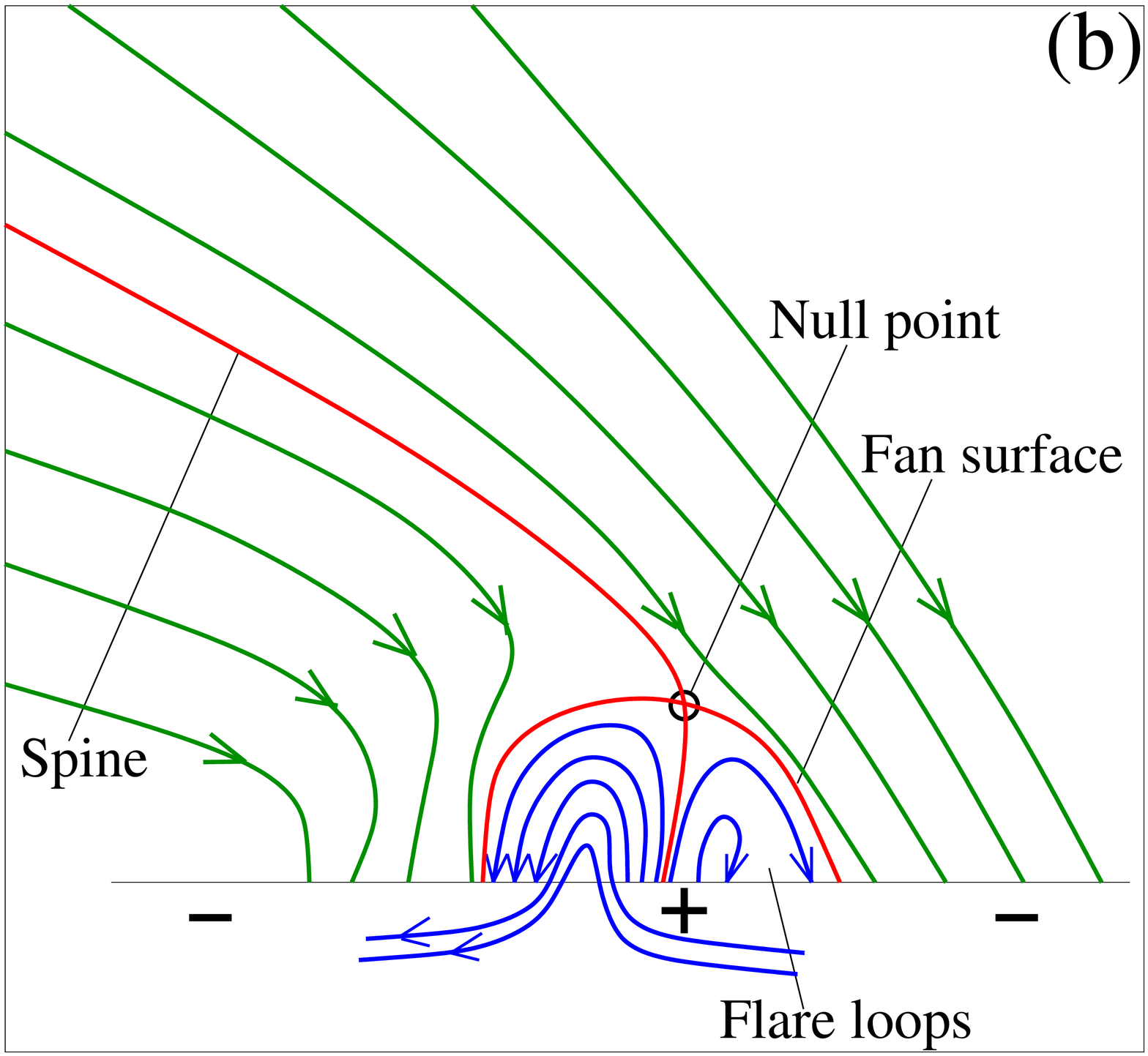}
 \plotone{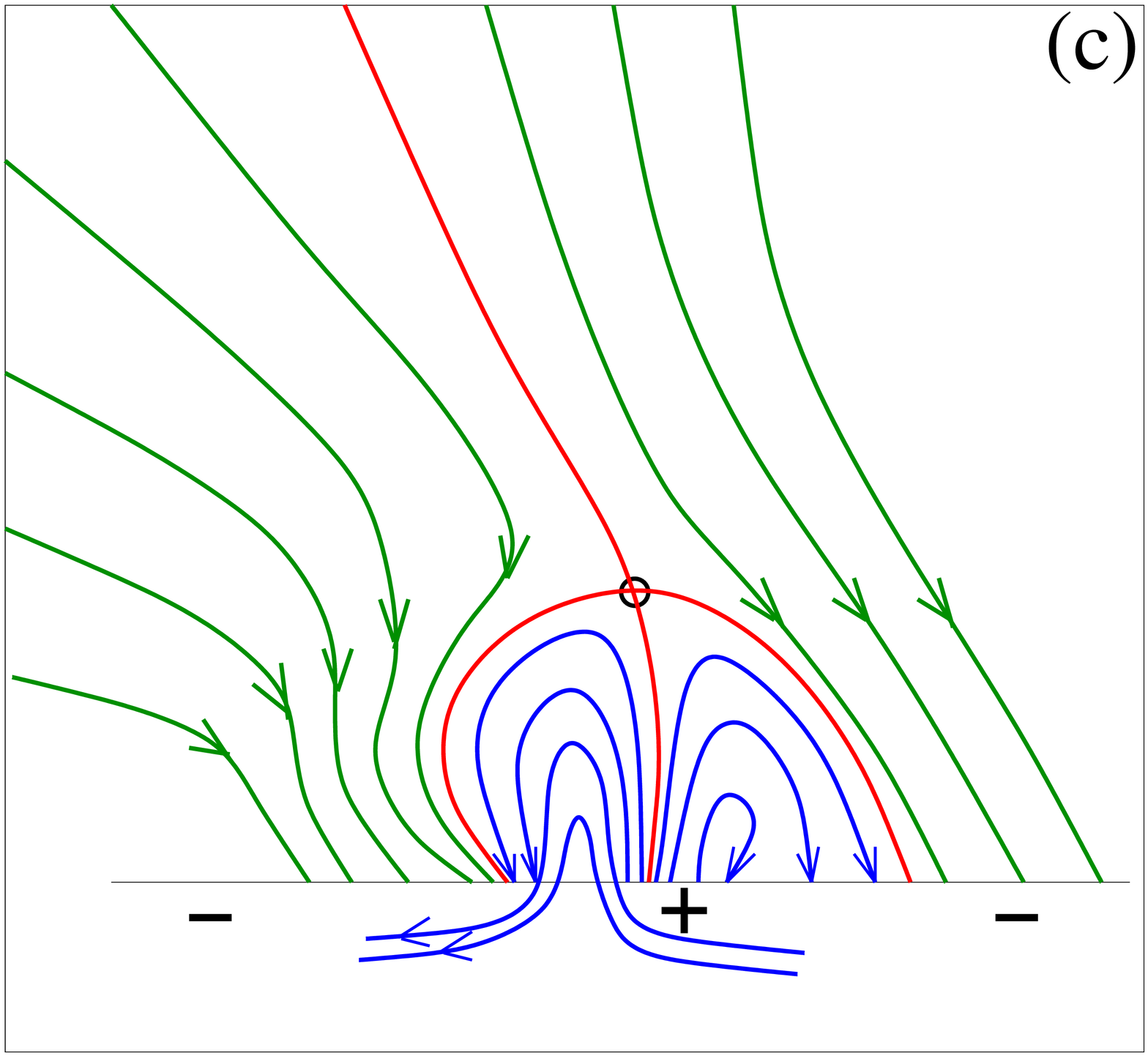}
 \plotone{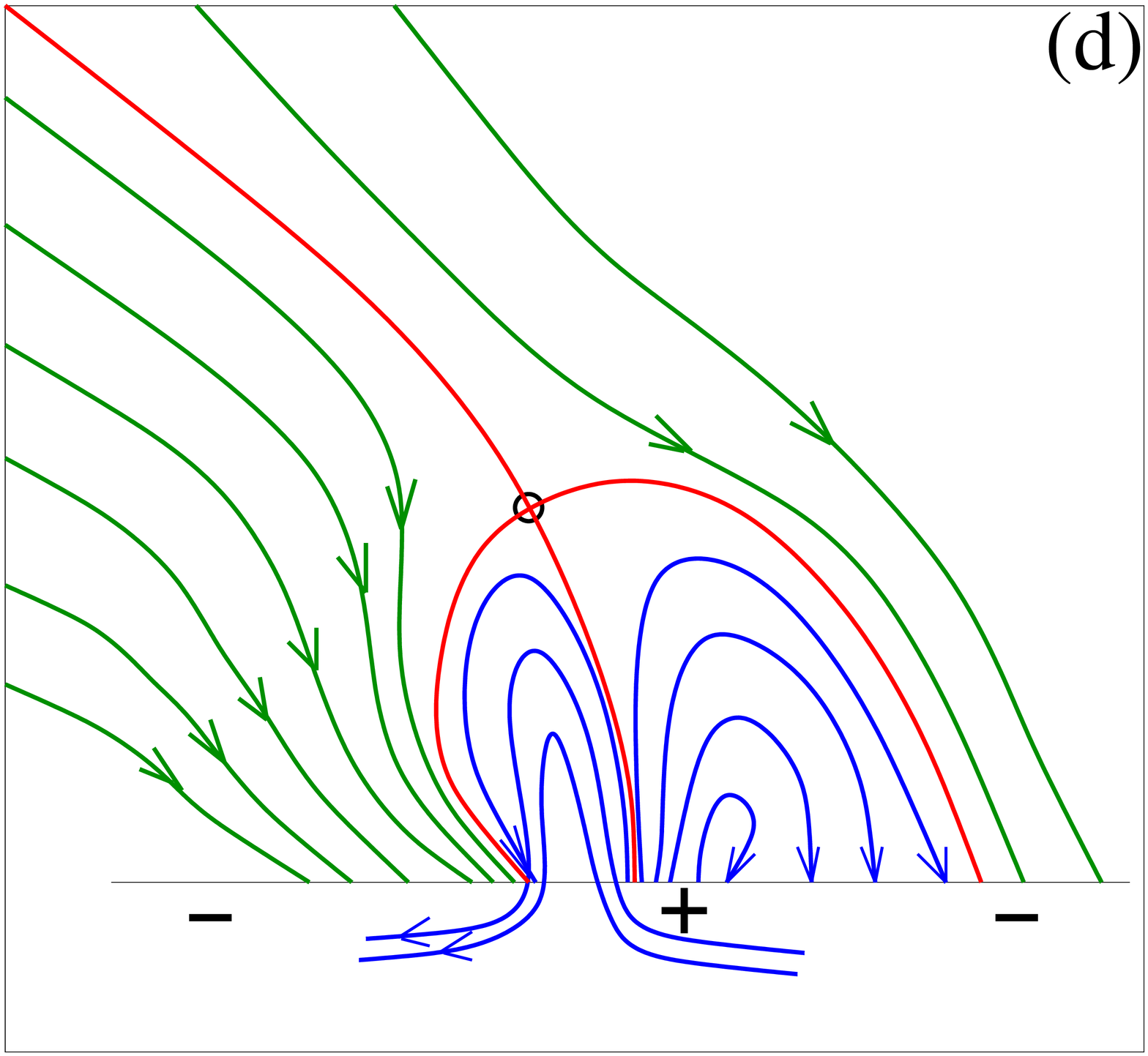}
 \caption[]{A 2D cut of the proposed model for the event development, indicating
 magnetic reconnection between the emerging flux and ambient field.
 The green field lines represent the unipolar surrounding field, the blue lines represent the closed
 field underneath the separatrix fan surface (dome), and the red lines represent the
 dome and spine. Beneath the dome, the emerging flux is located to the left of the inner spine 
 and the newly reconnected flux is to the right.
 Note the preferentially northward (toward the left) expansion of the fan surface,
 the changing orientation of the spine, and the migration of the null point
 from one magnetic flux surface to another as the result of reconnection.
 This is a simplified 2D rendition of the true 3D configuration with topological quasi-axisymmetry about the spine 
 (as shown in Figure~\ref{fan-spine.eps}(b)) and can be readily compared with our 2D observations.
 In a 3D case \citep[e.g., see figures in][]{PariatE.twist-jet-MHD.2009ApJ...691...61P, PariatE.twist-jet-homologous.2010ApJ...714.1762P}, 
 there are twists in the field and both sides of the spine are within a single topologically connected domain.
 } \label{cartoon-evolve.eps}	
 \end{figure*}
To examine the shape of the loop, we define the aspect ratio of the loop apex height 
to the half separation of the two legs in the north-south direction, 
 $ R \equiv h_{\rm loop} / [(y_{\rm NLeg} - y_{\rm SLeg})/2 ]$.
As shown in Figure~\ref{jet_vs_time.eps}(g), this ratio has a mean of $2.2>1$, indicating vertical elongation.
The elongation is reduced with $R$ decreasing from $\sim$3.2 to 1.8 	
during the gradual expansion phase (02:44~UT--$t_1$).
The loop then preserves its shape with a constant aspect ratio $\sim$1.8
during the common acceleration phase ($t_1$--$t_2$), but the elongation slightly increases	
afterward because of the earlier cessation of the lateral expansion.	
\begin{table}[bthp]	
\scriptsize	
\caption{Fitted velocities and accelerations of blobs in the northern leg of the loop at its late stage}
\tabcolsep 0.05in	
\begin{tabular}{crcrr}
\tableline \tableline

trajectory &  \multicolumn{1}{c} {linear fit}       &&  \multicolumn{2}{c} {parabolic fit} \\
               \cline{2-2}   \cline{4-5} 
or blob ID &  $\langle v \rangle$ ($\km \ps$)   &&  $a/g_{\sun}$  &   $v_{\rm final}$  ($\km \ps$)   \\			
\tableline 
A          &  $-94 \pm 5$   && ... & ... \\
B          &  $121 \pm 5$   && $11.4 \pm 3.1$ & $192 \pm 18 $  \\
C          &  $58 \pm 5$    && $11.1 \pm 2.6$ & $130 \pm 18 $   \\
D          &  $-95 \pm 4$  && ... & ... \\
E          &  $-57 \pm 2$   && $-4.8 \pm 0.7$ & $-115 \pm 9 $   \\

\tableline  \end{tabular}


\label{table_NLeg} \end{table} 

Near the end of the visible lifetime of the accompanying loop, the northern leg exhibits interesting behaviors.
There are two branches at this leg which initially expand upward during the loop growth.
When the elbow of the material bundle appears at $t_4$=02:51:44~UT as noted above, 
the two branches start to bend and {\bf retreat}
downward. We identified five bright blobs (A--E) as shown in Figure~\ref{mosaic_NLeg.eps},
among which blobs A and D represent the visible apexes of the two branches.
We then tracked the blob locations with time (Figure~\ref{NLeg.eps}(a)) and used blob sizes as uncertainties.
For each blob, its positions are well represented by a straight line fit, which gives its main direction of motion.
We obtained the distance along this direction and inferred the corresponding velocity and/or acceleration 
(Figures~\ref{NLeg.eps}(b) and (c)).	

We find that both blobs~A and D exhibit a downward acceleration followed by a deceleration within $\sim$1 minute,
and their average velocities are nearly $-100 \km \ps$ (see Table~\ref{table_NLeg}).
Near the onset (02:52:24~UT) of the northward sweep of the material bundle, 
blob~A becomes too vague to be identified, while two new blobs, B and C, appear nearby. 
They shoot upward with an acceleration $a \sim 11 g_{\sun}$ (where $g_{\sun}= 0.274 \km \pss$ is the
solar gravitational constant) and reach final velocities of $192 \pm 18$ and  $130 \pm 18 \km \ps$.
In contrast, as blob~D slows down its downfall near $t_5=$~02:52:40~UT, it turns its horizontal direction
from the north to the south (Figure~\ref{NLeg.eps}(a)) and we assign it a new blob ID named E.
This blob resumes a downfall with an acceleration of $a = (-4.8 \pm 0.7) g_{\sun}$ and reaches
a velocity of $-115 \pm 9 \kmps$ just before it plunges into spicules in the chromosphere.
Since the blobs are tracked by their projected 2D positions in the sky plane, these velocities
and accelerations are {\it lower} limits of their true values in 3D space. 
Meanwhile, the solar gravitational acceleration $g_{\sun}$ is the {\it upper} limit of its component along the unknown 3D trajectory.
Thus blob~E's downward acceleration is at least 5-fold greater than a corresponding free-fall
(dotted line in Figures~\ref{NLeg.eps}(b)).
Its final velocity is also 2--3 times larger than those (30--50~$\kmps$) found at the footpoints
of \Ha arch filaments \citep{BruzekA.arch-filam-drain.1969SoPh....8...29B, Roberts.arch-filam-drain.1970PhDT........19R}.
Even if a free-fall starts at a higher altitude at the top of trajectory~D,
it would reach a final velocity of only $42 \km \ps$, about ${1 \over 3}$ of trajectory~E's value.

\subsection{Streamlines of Falling Material: Inverted-Y Geometry}	
\label{subsect_invtY}

As mentioned earlier, jet material bound by gravity falls back to the chromosphere
and this continues throughout the duration of this event. The trajectories of the falling
material are smooth streamlines, which, at altitudes $ \gtrsim 20 \arcsec$,
are almost straight lines in the original direction of ascent. Kinematics of the falling material above
this height was investigated in \href{http://adsabs.harvard.edu/abs/2009ApJ...707L..37L}{Paper~I} using space-time plots (see Figure~2 there). 
Below this height, the streamlines are curved as if they bypass an unseen dome or null point.
This can be clearly seen in the online Animation~1, especially from 03:25 to 03:35~UT.

To highlight these streamlines, we performed running difference
between each pair of images 2~minutes apart (every 15 frames at an 8~s cadence) within
a selected duration, and superimposed the positive parts of all differenced images.
A sample of the results is shown in Figures~\ref{streamline.eps}(a)--(c).
We then visually traced the streamlines, a collection of which clearly outlines an inverted-Y geometry
(Figure~\ref{streamline.eps}(d)).
By overlaying these streamlines on top of multiwavelength images (Figures~\ref{streamline.eps}(e)--(h)), 
we note the following spatial relationships: 
 (1) some streamlines, particularly those on the right-hand side, 
are distributed close to the growing Ca loop at its final stage;
 (2) the inverted-Y goes around the cusp-shaped flare loop seen in SXR;
and (3) three branches (labeled $b_0$, $c_1$, and $a_1$) of the streamlines are cospatial with
the dark absorption features seen in EUV, which likely represent the same cool material that emits in \CaIIHA.
We also note that 	
the drift of the material bundle's lower end terminates at 02:54~UT
near the bifurcation of the streamlines 1~hr later		
(Figures~\ref{mosaic.eps}(l) and \ref{streamline.eps}(c)).

\section{Interpretation of Observations}
\label{sect_discuss}


The observations presented above bear two significant implications.
First, the Ca loop comes into sight on the visible side of the limb, 
starts to grow into the chromosphere among neighboring spicules, and makes its way into the corona, 
expanding vertically and laterally. This suggests that the growing loop results from the {\bf emergence}
of a magnetic bipole from below the photosphere. 
This inference is supported by the following three factors:
 (1) The loop's initial upward expansion 
velocity of $16.2 \pm 0.4 \kmps$ 	
is of the same order of magnitude as the 10--15 $\kmps$ Doppler velocities of rising arch filaments 
in EFRs \citep{ChouZirin.arch-filament-rise.1988ApJ...333..420C}, 
as well as the $\sim$$20 \kmps$ speeds of emerging flux ropes found
in 3D MHD simulations \citep{Archontis.emergeMHD.2004A&A...426.1047A, 
FanY.rot-sunspot.2009ApJ...697.1529F, Martinez-Sykora_MHDemergenceII.2009ApJ...702..129M}
and of recently discovered prominence plumes \citep{BergerT.promin-plume.2008ApJ...676L..89B}.
 (2) The separation ($\lesssim$15$\arcsec$) of the flare kernels (at loop footpoints)
is only $1 \over 2$ of the separation between the Ca loop legs, which is
expected for emerging flux due to its rapid lateral expansion in the corona.
 (3) In addition, it is possible that the earlier, weaker surge%
  \footnote{This earlier surge and the main jet, together with another jet occurring 10 hrs 
  later \citep{Nishizuka.giantCaHjet.2008ApJ...683L..83N}, could be part of the recurring jet activity,
  which has been observed in other events \citep{ChiforC.Hinode-jet2.2008A&A...491..279C}
  and simulated with MHD models 
  \citep{Archontis.recurring-jets.2010A&A...512L...2A, PariatE.twist-jet-homologous.2010ApJ...714.1762P}.
  }
(02:14--02:30~UT) is the initial signature of flux emergence. If so,
the 30~minutes delay of the first appearance of the Ca loop at 02:44~UT is consistent with the time scale
for the emergence of a typical small-scale ephemeral region.
This time is required for the emerging flux to be built up in order to generate sufficient upward pressure force 
and to drain heavy material carried from the interior before it can further expand into the corona.

We note that, other than flux emergence, footpoint shearing, twisting, or braiding motions of a {\it pre-existing} 
coronal structure, such as an arcade of coronal loops,
can also increase its magnetic stress and helicity and cause it to expand or even erupt
\citep[e.g.,][]{AntiochosS.breakout2.1999ApJ...510..485A, Rachmeler.MHD-jet.2010ApJ...715.1556R}.
So we examined available data from 	
\hinodeA, \traceA, and \stereoA.	
However, we found no indication (either emission or absorption) of the prior existence of the loop or material bundle
at any size resolved by the instruments (down to $0\farcs 2$) and
up to two hours before this event.	
These data include multiple wavelengths --- 	
\ion{Ca}{2}~H, EUV (171, 195, 284, and 304~\AA), and soft X-rays --- which cover a wide range of temperatures ($10^4$--$10^7 \K$).
This essentially rules out the possibility that the appearance of the Ca loop results from temperature changes
and/or expansion of a pre-existing coronal structure. 
Hence, flux emergence still remains the most likely possibility for
the origin of the Ca loop, although we have no reliable magnetograms of this near-limb region
to provide direct evidence.

The second implication of our observations is that, as mentioned in Section~\ref{sect_intro}, 
any bipole emerging into a unipolar region, 
i.e., the coronal hole in this case, can naturally 
lead to a {\bf fan-spine} topology 
\citep{AntiochosS.breakout.1998ApJ...502L.181A, Torok.fan-spine.twist-emerg.2009ApJ...704..485T}.
The streamlines of the falling material, assumed to be parallel to magnetic
field lines, clearly indicate such a configuration
as shown in Figure~\ref{fan-spine.eps}. 	
In particular, streamlines $a_0$ and $b_0$ evidently bifurcate, 
suggesting that the spine is located between them and the null point lies close to the bifurcation. 
The fact that the streamlines avoid the dome indicates that the jet material
originates from above the separatrix fan surface.
There are, however, outliers to this general trend, and streamlines $b_1$ and $c_1$ 
apparently pass through the dome. This is likely a projection effect of the true 3D geometry,
in which these streamlines lie behind or in front of the dome that has a finite extent along 
the line of sight (e.g., Figure~\ref{fan-spine.eps}(b)). 

Note that the 3D fan-spine configuration has a variant form in a 2D geometry with translational
invariance in the third dimension or in a 3D geometry with a significantly elongated parasitic polarity.
In this case, the {\it 3D null point} is replaced with a {\it separator line} and the spine line is replaced
with a spine surface. This surface divides the volume underneath the dome into two topologically separate 
chambers \citep[see Figure~3 of][]{Moreno-Insertis.EISjet2008ApJ...673L.211M},
which are, however, a single connectivity domain in the true 3D null case. 
The observations of flow streamlines very likely located behind or in front of the 
inferred dome provide a basis for our adoption of the true 3D null 
model for our discussion in the rest of the paper.




\subsection{Proposed Model}
\label{subsect_model}



Based on the above two inferences, we postulate that the earlier {\it material bundle}
and later {\it collimated jet}	
represent the {\it outer spine} and its neighboring field lines in different stages, while the
{\it growing Ca loop} is a 2D rendering (projection) of the entire 3D {\it dome or separatrix fan surface}
(Figure~\ref{fan-spine.eps}). We believe that, among alternative models, this conjecture best explains
the observations. The postulated event develops as follows (see Figure~\ref{cartoon-evolve.eps}).

When a twisted flux rope emerges into the corona in an open-field environment, 
the flux rope expands rapidly, driven by its considerable magnetic pressure,
and presses onto the ambient field to form a current sheet at the discontinuity between them
\citep{HeyvaertsJ.flux-emerg-flare-model.1977ApJ...216..123H}. 
Magnetic reconnection ensues and results in a fan-spine structure mentioned above. 
The outer spine originates from the null point and opens to infinity, while the inner spine 
apparently divides the vault under the dome into two parts seen in 2D projection. 

Reconnection outflows continuously pump dense	
plasma upward along the newly reconnected field lines near the {\bf outer spine}.
Emission from such plasma (hatched region around the spine in Figure~\ref{fan-spine.eps}(a))	
could be responsible for the observed {\bf material bundle} and later {\bf collimated jet},
whose diffuse upper end and relatively sharp lower end
(Figures~\ref{mosaic.eps}(f)--(h)) are readily explained by the geometry of the outer spine.
Reconnection also transfers magnetic helicity from the emerging flux rope into the newly reconnected
open field. This and plasma pumping are both manifested in the upward swirling motion of
material along the helical trajectory observed around 02:47~UT (Figures~\ref{mosaic.eps}(c)--(d)).
Under the influence of the Lorenz force, the twists in the reconnected open field 
tend to unwind themselves and drive these field lines to rotate, 
as found in spinning \Ha surges 
\citep{Shibata.Uchida.helic-jet.1986SoPh..103..299S, Canfield.surge-jet1996ApJ...464.1016C}.
Such spins propagate up toward the open end in the form of torsional MHD	
waves, and their 2D projections would
appear as traveling sinusoidal oscillations, just as observed here (Section~\ref{subsect_bundle}).
The inferred rotational velocities and periods (Figure~\ref{perp_fit.eps}) are comparable to
those of transverse oscillations found in X-ray jets \citep{Cirtain.XRTjet2007Sci...318.1580C},
prominences \citep{Okamoto.Alfven-wave.prominence.2007Sci...318.1577O},
and coronal loops \citep{Ofman.Wang.SOT-wave.2008A&A...482L...9O}, 
which were interpreted as signatures of Alfv\'{e}n or fast kink mode waves.
Our inferred phase speed of $v_{\rm ph} =786 \pm 30 \kmps$ is also within the range of
typical coronal Alfv\'{e}n or fast-mode speeds.
At the same time, the Lorentz force associated with the twists and axial gradient of currents
\citep[e.g., see the current density in Figure~9 of][]{PariatE.twist-jet-MHD.2009ApJ...691...61P}
may have a strong axial component that can drive the upward ejection of material along 
the axis of the bundle \citep{BellanP.plasma-lab-jet.2005Ap&SS.298..203B}.

As the dense emerging flux rope expands and sweeps through the dense lower atmosphere ahead of it, 
a layer of enhanced local density is expected to form at its leading front.
In our case, this layer is the {\bf fan surface} between the emerging flux and the ambient open
field, marked as the hatched region on the dome in Figure~\ref{fan-spine.eps}(a)).
In 2D projection, it could appear as the {\bf growing Ca loop} seen here.
The expansion rate of the flux rope, presumably depending on the flux emergence rate 
and the interplay between the strengths of the emerging field and the surrounding field,
controls the dynamic evolution. Early in the event (before $t_1$=02:49:02~UT), 
the flux rope (thus the fan surface) expands at a moderate speed ($16.2 \pm 0.4 \km \ps$),
giving the ambient field a {\it gentle} push. This leads to moderate rates of reconnection 
and supply of mass and helicity to the {\it material bundle} around the spine, which evolves in a quasi-static manner.
As sufficient twists have emerged into the corona with time, a kink-like instability can occur
and force the flux rope to undergo an accelerating expansion 
\citep[][]{FanGibson.emrgKink.popular.2004ApJ...609.1123F,
PariatE.twist-jet-MHD.2009ApJ...691...61P, Rachmeler.MHD-jet.2010ApJ...715.1556R}, 
{\it strongly} pushing the ambient field and thus driving rapid reconnection. 
If we interpret flares to be indicators of rapid energy release by fast reconnection, 
we may identify time $t_1$ as the onset of the observed flare heating in this event.
In addition, the increased fluxes of mass and helicity transferred to the material bundle through reconnection
can no longer relax in a quasi-static manner, eventually resulting in a runaway instability
that leads to the {\it collimated jet}.
This explains the simultaneous transitions from {\bf slow to fast evolution} 
for both the material bundle and growing loop (Figure~\ref{jet_vs_time.eps}).
This is also energetically analogous to the two regimes found by 
\citep{PariatE.twist-jet-homologous.2010ApJ...714.1762P}:
slow, quasi-static reconnection in the energy-storage phase 
and fast, dynamic reconnection in the energy-release phase in a rotating current sheet
associated with the kinking flux under the fan surface.

Because of continuous reconnection, the material bundle around the spine is a 
constantly {\bf evolving entity}.	
A given field line in the bundle is illuminated only when reconnection 
drives 	
dense mass flow along it. It becomes invisible once the temperature leaves the
\CaIIH response range ($\lesssim$$2\E{4} \K$) or the density drops significantly, which could be
true at high altitudes or when the reconnection site migrates away. 
 In the {\it former} case, the material bundle would correspond to the lower portion of a jet or surge,
whose upper portion has upflows at reduced density and brightness. This gives rise to the diffuse appearance of the 
upper end of the bundle. This also explains why the streamlines of the falling material are distributed in
a large volume and in various directions (Figure~\ref{streamline.eps}), not just along the path of the final 
collimated jet. This is because material is continuously ejected upward in different directions
when the bundle swings, as what happens when a fireman swings his firehose.
 In the {\it latter} case of reconnection site migration, the visible threads in the material bundle 
are constantly being replaced and we see different field lines over time as reconnection develops. 
This could explain why each sinusoidal oscillation (Figure~\ref{tslice_perp.eps}) appears ephemerally.

As the system seeks its lowest energy state, the open-ended spine field line can 
change its orientation, while the null point migrates in space from field line to field line, 
depending on the dynamic evolution of reconnection (Figure~\ref{cartoon-evolve.eps}). 
This seems to reflect what we observe here and explain the elbow  
(Figure~\ref{mosaic.eps}) and the drift of the material bundle's lower end 
(Figure~\ref{streamline.eps}(c)). The latter is likely located near the null point
and its overall upward migration may result from the upward development of reconnection
as predicted in the standard flare model \citep{SturrockP1966Natur.211..695S, KoppR1976SoPh...50...85K}
or from a cumulative effect of advection motions driven by gas pressure in 
the neighborhood of the null where the Lorentz force vanishes \citep{PariatE.twist-jet-MHD.2009ApJ...691...61P}.
The low angle of the material bundle in its early stage probably reflects the strong
horizontal field at low altitudes. 
The back-and-forth fast ($151 \kmps$, see Figure~4 in \href{http://adsabs.harvard.edu/abs/2009ApJ...707L..37L}{Paper~I}) 
swing of the entire material bundle possibly results from
catastrophic release of excessive twists into the open field, similar to the kink or writhe of a flux rope.
Also, if we assume the measured twist propagation velocity $786 \kmps$ to be the \Alfven velocity $v_A$,
the velocity ratio $151/786=0.19$ is close to the 0.2 value of the drifting outer spine
found in the 3D simulation of \citet{PariatE.twist-jet-MHD.2009ApJ...691...61P}.

\subsection{Discussion}
\label{subsect_discuss}

Many characteristics of our observations and the above interpretation resemble
those in the simulation of \citet{Torok.fan-spine.twist-emerg.2009ApJ...704..485T},
particularly their fan-spine topology with a 3D null point resulting
from emergence of a bipole into a locally unipolar region and the launch of a torsional MHD wave
from the reconnection site.
However, in their case, magnetic reconnection develops in two steps and each step leads to 
the formation of hot loops on each side of the spine, which constitute one half of a full anemone,
consistent of the \hinode XRT observations of a specific jet event. 
In our case, we found no signature of such two-step reconnection, 
and we could not identify the expected inner spine emission 
(hatched with dashed lines in Figure~\ref{fan-spine.eps}(a)) 
in available data, presumably due to unfavorable temperatures and/or densities under the dome. 
However, the flare loop in SXR and EUV (Figures~\ref{streamline.eps}(g)--(h)) and its right branch in 
\CaIIH (Figure~\ref{sot_trace_xrt.eps}(j)) are located in the expected position
to the right hand side of the null point for the newly reconnected field.
We note that the upward expansion of the emerging flux toward the north has a large component {\it parallel}
to the ambient field and can readily makes its way in a herniation manner, 
while the expansion toward the south must {\it perpendicularly} press the ambient field against strong resistance.
This topological {\bf asymmetry} can therefore preferentially facilitate
the northward expansion, which is at least 4 times faster than its southward counterpart 
(Figure~\ref{jet_vs_time.eps}(f)).
This is similar to the simulation of the primarily vertical (vs.~horizontal) expansion of the dome 
in a vertical ambient field \citep{PariatE.twist-jet-MHD.2009ApJ...691...61P}.
In addition, the null point can be more readily advected in the spine direction, as manifested
in its overall upward and northward migration (Figure~\ref{streamline.eps}(c)).

The northern leg starts to reverse its expansion and {\bf retreat downward} at $t_4=$~02:51:44~UT. 
In addition, when the material bundle sweeps toward the north, its lower end, assumed to be near the null point, 
slightly drops in height by $3 \arcsec$ (Figures~\ref{jet_vs_time.eps}(d)).	
These occur near the peak of the flare, possibly when a significant amount of energy has been released.	
This is consistent with the implosion conjecture of \citet{HudsonH.implosion.2000ApJ...531L..75H}
and its theoretical demonstrations 
\citep{Zhang.Low.helicity.conserv.2003ApJ...584..479Z, Janse.Low.implosion.2007A&A...472..957J}.
It predicts that after major energy release, the magnetic pressure (or energy density) in the local volume 
is significantly reduced and the resulting imbalance with the pressure in the surroundings would push 	
this volume to contract. Observational evidence of this conjecture has gradually emerged 
\citep{LiuR.implosion.2009ApJ...696..121L, LiuW_FPAsym_2009ApJ...693..847L}.
When this happens, opposite to an expanding twisted flux rope in which twists tend to accumulate 
in the expanded top portion \citep[][p.~189]{ParkerE.cosmicB-book.1979cmft.book.....P},
more twists will be pushed by the Lorentz force to concentrate in the leg portion when the rope contracts. 
This explains why twists become more visible in the northern leg later when it contracts.
This downward Lorentz force might also provide additional push to the material and help explain the
faster than free-fall {\it downflow} observed here (Figures~\ref{mosaic_NLeg.eps} and \ref{NLeg.eps}).

At the same time, some material is {\bf ejected upward} at accelerations up to $(11.4 \pm 3.1) g_\sun$ 
(Figures~\ref{mosaic_NLeg.eps} and \ref{NLeg.eps}). It is quite puzzling for this to take place
simultaneously in the same northern leg as the fast downflow, even though they might occur 
on different field lines at separate line-of-sight positions.	
In any case, there are several possible explanations which we cannot distinguish with the available data:
(1) If the ejection is on twisted field lines that reconnect with untwisted open field lines, it could be
driven by the Lorentz force associated with the upward {\it relaxation} of twists, which seems to happen here 
(Figures~\ref{mosaic_NLeg.eps}(g)--(o)). 
(2) If the upward ejection and downflow are indeed on the same field lines, they could be the
upward and downward branches of the secondary outflows bifurcated from the primary reconnection
outflow as predicted in MHD simulations 
\citep{YokoyamShibata.jetModel1996PASJ...48..353Y, Moreno-Insertis.EISjet2008ApJ...673L.211M}.
(3) Another less likely possibility is that the ejection and downflow are oppositely directed
outflows from reconnection occurring at an X-point between them. An X-type null point at this location is, however,
not favored by our adoption of the spine-fan topology to interpret this event.


The {\bf disappearance of the \CaIIH loop} possibly results from the combination of several processes,
including mass drainage, temperature variation, and topological change.
The first two mechanisms are expected to operate gradually, while the third one could
happen catastrophically. 
  First, as a flux rope emerges and plows through the lower atmospheres, 	
dense material is dredged into the hot corona. Part of this material	
would contribute to the Ca emission seen at the fan surface.
Such dense material is expected to be pulled back by {\bf gravity} and slide down the dome,	
as seen in \Ha arch filaments \citep[e.g.][]{ChouZirin.arch-filament-rise.1988ApJ...333..420C}.
This is because the scale height of the $\lesssim$$2\E{4} \K$ plasma emitting the \ion{Ca}{2}~H line
is only $\lesssim$1.2~Mm and pressure gradient from gravitational stratification is simply too small 
to support the weight of plasmas extending 	
to the height of the dome ($>$10~Mm).	
In our case, such mass loss is evident in the northern leg of the loop.		
This would make the apex of the dome have the lowest density
and become less well-defined and then invisible first, 
before the loop legs gradually fade away (Figures~\ref{jet_vs_time.eps}(e) and (f)).
  Second, {\bf heating} of the Ca loop (fan surface) may occur during the course of its rise, 
possibly as a result of reconnection with the ambient field. 
Once its temperature rises above the $\lesssim$$2\E{4} \K$ range,
the \CaIIH loop would disappear. This can be seen in Figure~\ref{sot_trace_xrt.eps}(h), where the northern
leg of the loop is vague in Ca but prominent and bright at 195 \AA. Later, because of significant mass drainage that has considerably
reduced the density of the loop (fan surface), it would no longer appear as detectable absorption or emission
in EUV or SXR, as we see here (Figures~\ref{sot_trace_xrt.eps}(j)--(k)). 
  Finally and more importantly, a catastrophic topological change, which may be involved in the launch
of the chromospheric jet, can alter field connectivity and contribute to the disappearance of the 
fan surface in emission.

We note in passing that there is an {\bf overarching loop brightening in SXR} at 02:44~UT (see Figure~\ref{sot_trace_xrt.eps}(b))
that appears simultaneously with the bright \CaIIH loop (dark in EUV) but $\sim$$10 \arcsec$ higher in altitude.
The southern leg of the Ca loop appears as dark absorption in SXR (Figure~\ref{sot_trace_xrt.eps}(f)),
suggesting that the SXR loop brightening is located behind the Ca loop.
When the legs of the Ca loop disappear around 02:55~UT (Figure~\ref{jet_vs_time.eps}(f)),
this SXR loop becomes invisible too (Figure~\ref{sot_trace_xrt.eps}(j)). 
These timing coincidences seem to suggest that the SXR loop is of the same emerging flux system as the
Ca loop but at higher altitudes and temperatures. However, during its lifetime, the SXR loop hardly changes
its size or shape, while the Ca loop has risen more than $20 \arcsec$, well beyond the XRT's $2 \arcsec$ spatial resolution. 
This contradicts the expectation that the overarching SXR loop would expand together with the low lying Ca loop if they were of the same
emerging flux system. Another possibility is that the SXR loop brightening represents heating
during magnetic reconnection in a curved current sheet between the emerging and ambient fields,
as suggested by \citet{Yoshimura.AFS-SXR-loop-relation.1999ApJ...517..964Y} who found
SXR brightening above emerging \Ha arch filaments. However, the location of the event within
the inferred coronal hole poses a challenge for the existence of such a hot SXR loop as observed here.
The nature of this transient SXR loop brightening thus remains an open question.

\section{Conclusions}	
\label{sect_conclude}


We have presented multiwavelength observations and detailed analysis of a chromospheric jet 
and its accompanying growing loop. This extends the study presented in \href{http://adsabs.harvard.edu/abs/2009ApJ...707L..37L}{Paper~I} which focused on the fine 
structure and kinematics of the jet itself. We summarize our new observations as follows.

\begin{enumerate}

\item	
Potential field extrapolation indicates that this event occurs in an equatorial {\bf coronal hole} and 
as expected, the jet is closely aligned with the open field lines	
(Figures~\ref{pfss_global.eps} and \ref{sot_Bfield.eps}).

\item	
The predecessor of the jet is a {\bf bundle of material threads} ($\lesssim$$1 \arcsec$ wide)	
extending from the chromosphere into the corona. This bundle exhibits transverse
{\bf sinusoidal oscillations} across its axis, whose velocities range from $47 \pm 9$ to $58 \pm 11 \kmps$,
periods from $162 \pm 11$ to $197 \pm 35 \s$, and amplitudes from $1.5 \pm 0.3$ to $1.7 \pm 0.3 \Mm$.
Such oscillations propagate upward at velocities as high as $v_{\rm ph} =786 \pm 30 \kmps$
(Figures~\ref{tslice_perp.eps} and \ref{perp_fit.eps}).
We interpret these as evidence of propagating torsional MHD		
waves. 

\item	
The material bundle first slowly and then rapidly swings up, with the orientation angle of its central axis from the limb 
growing by $>$$50\degree$ in 10 minutes (Figure~\ref{jet_vs_time.eps}). The transition from the {\bf slow to fast swing} 
phase coincides with the onset of an A4.9 flare, which heats the plasma to $T= 12.2 \pm 0.6 \MK$. 
The bundle then swings back in a whiplike manner and develops into
a collimated jet (Figure~\ref{mosaic.eps}), which continues to exhibit transverse oscillations 
(see \href{http://adsabs.harvard.edu/abs/2009ApJ...707L..37L}{Paper~I}), but at fractionally slower rates than the earlier bundle mentioned above.

\item	
A {\bf loop expands} simultaneously in these two phases. It attains a uniform vertical velocity 
of $16.2 \pm 0.4 \kmps$ during the {\it gradual phase} and reaches $135 \pm 4 \kmps$ at the end
of the {\it acceleration phase} (Figure~\ref{jet_vs_time.eps}). The initial slow rise velocity is similar to
those of emerging fluxes found in \Ha arch filaments and in MHD simulations. 
The lateral expansion is asymmetric and dominated by the northward displacement of the northern leg. 

\item	
The loop appears to rupture or collapse near the peak of the flare and its apex becomes undetectable first. 
The northern leg of the loop {\bf retreats downward} and material drains
down to the photosphere at accelerations ($a= (-4.8 \pm 0.7) g_\sun$) greater than free-fall. 
At the same time, some material is {\bf ejected upward} ($a=(11.4 \pm 3.1) g_\sun$) in the same leg 
(Figures~\ref{mosaic_NLeg.eps} and \ref{NLeg.eps}; Section~\ref{subsect_discuss}).

\item	
Some material falls back along streamlines in the original direction of ascent, showing 
no more transverse oscillations (\href{http://adsabs.harvard.edu/abs/2009ApJ...707L..37L}{Paper~I}). Most of the streamlines 
swerve around an inferred dome extending above the chromosphere and characterized with a null point at its top,
depicting an {\bf inverted-Y} geometry.
These streamlines closely match in space the 
late Ca loop prior to its rupture, the X-ray flare loop, and the 
EUV absorption features (Figure~\ref{streamline.eps}).	

\end{enumerate}


We interpret (Section~\ref{sect_discuss}) these observations in the framework of
the emergence of a twisted flux rope into 
an open field environment leading to the formation of a jet through magnetic reconnection 
\citep[e.g.,][]{HeyvaertsJ.flux-emerg-flare-model.1977ApJ...216..123H,
YokoyamShibata.jetModel1995Natur.375...42Y, Moreno-Insertis.EISjet2008ApJ...673L.211M},
and the relaxation of twists transferred into the jet leading to its spin 
\citep{Shibata.Uchida.helic-jetMHD.1985PASJ...37...31S, Shibata.Uchida.helic-jet.1986SoPh..103..299S, 
Canfield.surge-jet1996ApJ...464.1016C, Torok.fan-spine.twist-emerg.2009ApJ...704..485T}.
We further identify signatures of the {\bf fan-spine} topology throughout the event, 
from the precursor to post-eruption evolution.
The {\it outer spine} is recognized as the {\it material bundle} that eventually develops into the {\it collimated jet},
while the {\it fan surface} is imaged as the {\it growing Ca loop} in projection.
After the eruption, the presence of this magnetic skeleton is clearly implied by the streamline geometry of
the falling material.


	%
	%
	%

Our observations and model share commonalities with their counterparts in the literature,
while the major differences given by our new findings are as follows.

\begin{enumerate}

\item	
The {\bf simultaneous} growth of the emerging flux and development of the resulting jet, 
synchronized in two stages (Figure~\ref{jet_vs_time.eps}), have been clearly established here
for the first time, to the best of our knowledge. Growing loops (other than post-flare loops) were recently noted
in X-ray jet events \citep{ShimojoM.XRT-fine-jet.2007PASJ...59S.745S, ChiforC.Hinode-jet2.2008A&A...491..279C},
but their detailed temporal evolution and relationship with the jet were not clear.

\item	
In previous models \citep[e.g.][]{YokoyamShibata.jetModel1995Natur.375...42Y}, reconnection between
the emerging flux and the overlying field would immediately lead to the launch of an eruptive jet. 
In our case, when the reconnection rate is {\it moderate} early in the event, the jet,
manifesting itself as the material bundle, undergoes {\it quasi-static} evolution for more than 20 minutes. 
We call this an ``{\bf intermediate jet}" stage, which later develops into the classical {\it eruptive} jet 
as a result of {\it fast} reconnection driven by the accelerating expansion of the emerging flux.
These are analogous to the slow and fast reconnections in the energy-storage and -release stages of coronal jet simulations, respectively \citep{PariatE.twist-jet-homologous.2010ApJ...714.1762P}.

\item	
The whiplike motion of a jet has been predicted as a consequence of the sling-shot effect 
of the newly reconnected field lines
\citep[e.g.,][]{Shibata.Uchida.helic-jetMHD.1985PASJ...37...31S, Canfield.surge-jet1996ApJ...464.1016C},
and has been observed as a {\it unidirectional} swing {\it away} from the accompanying flare where reconnection occurs.
In our case, the axis of the material bundle {\bf swings back and forth}, and
the previously predicted whip motion only applies to the second swing here when the material bundle
moves into its collimated jet position. We interpret this as instabilities possibly	
related to the catastrophic unload of excessive twists (Section~\ref{subsect_model}).

\end{enumerate}


A statistical study of similar \hinode events is required before more general conclusions can be drawn.
The validity of our phenomenological interpretation shall also be rigorously checked against
theoretical models, numerical simulations \citep[e.g.,][]{PariatE.twist-jet-homologous.2010ApJ...714.1762P}, 
and laboratory experiments \citep[e.g.,][]{BellanP.plasma-lab-jet.2005Ap&SS.298..203B}.
Finally, as pointed out by \citet{AntiochosS.breakout.1998ApJ...502L.181A}, the emergence of a bipolar flux 
system in a unipolar region on the photosphere naturally produces a 
local minority-polarity region with a spine-fan helmet magnetic 
structure above it.  Our observations clearly show such an event and have 
identified its observable dynamical characteristics, 
thus providing motivation for future investigation of the rich 3D magnetic topologies 
to be found in such structures.



\acknowledgments
{

This work was supported by \hinode SOT contract NNM07AA01C.		
Hinode is a Japanese mission developed and launched by ISAS/JAXA, with NAOJ as
domestic partner and NASA and STFC (UK) as international partners. It is operated
by these agencies in cooperation with ESA and NSC (Norway). 
We thank the anonymous referee for constructive comments and helpful suggestions.
We are grateful to Paul Bellan, Marc DeRosa, Mark Cheung, Juan Martinez-Sykora, Yu Liu,
Haimin Wang, Richard Shine, and Rita Ryutova for discussions and help of various kinds, 
to Yan Xu and Shuo Dai for help with YNAO \Ha data reduction,
to Mei Zhang for comments on the manuscript,
and to Dominic Zarro and William Thompson for help with \stereo maps.
}





\appendix

\section{A.~Image Processing and Cross-Instrument Coalignment}
\label{Append_sect_coalign}

We describe below the procedures to process and coalign images obtained from \hsiA, \traceA,
\hinode XRT and SOT, YNAO, and \stereo EUVI. 
The absolute solar coordinates of \hsi images have sub-arcsecond accuracy owing to
its limb sensing aspect system and star based roll angle measurements \citep{FivianM2002SoPh..210...87F}. 
We thus used \hsi images as a fiducial for our coalignment.

XRT images were processed using the \texttt{xrt\_prep} and \texttt{xrt\_jitter} routines.
We selected three XRT images (at 02:50:49, 02:51:51, and 02:53:51~UT, exposure $<1$~s) 
during the flare in which over-exposure was minimal. 
We then constructed three \hsi images at 3--6~keV with 100~s integration centered at these times.
Assuming these XRT and \hsi emissions were from the same source at each time,
we shifted the XRT image to match its centroid of the 99\% brightness contour
and that of the 50\% contour of the corresponding \hsi image. 
The averages of the required shifts at the three times	
are $\Delta x = 10 \farcs 3$ and $\Delta y = 11 \farcs 3$			
in the solar west and north directions, respectively.
By rotating the shifted XRT images about the \hsi centroid to match the limb 
in a trial and error manner, we found that no additional rotation is needed
and the $y$-axis is already the solar north.

To coalign \trace 195 \AA\ images, we noted that both \trace and XRT observed flare loops
except that XRT loops are hotter and located at slightly higher altitudes.
We thus selected a pointing-corrected XRT image at 02:53:51~UT and a later \trace image at 02:59:14~UT
(when the hot XRT loops had cooled down) to match their flare loop legs.
The \trace image was required to be shifted by
$\Delta x = -3 \arcsec$ and $\Delta y = 14 \arcsec$,		
and then rotated by $1 \degree$ counter-clockwise about the centroid of
the flare loop, $O=[975 \farcs 8, -42 \farcs 7]$, to match the limb.

For SOT observations, we first processed \ion{Ca}{2}~H images using the standard \texttt{fg\_prep} routine
with the \texttt{/no\_pointing} option to disable automatic pointing correction which is currently not satisfactory.
We then made relative coalignment by cross-correlating neighboring images. 
To find the absolute image coordinates, we noted that the $\lesssim$$2\E{4} \K$
\ion{Ca}{2}~H emitting plasma appeared as absorption in \trace 195 \AA\ images.
We then selected a pointing-corrected \trace image at 02:43:20 with a 65~s integration
and summed the corresponding SOT images from 02:43:23 to 02:44:19~UT.
By matching features including the jet material bundle, large spicules, and the limb, 
we found that the center of the SOT image must be placed at
$x=998 \farcs 9$ and $y=-32 \farcs 3$, and then the image should be
rotated by $0.2 \degree$ clockwise about point $O=[975 \farcs 8, -42 \farcs 7]$.

YNAO \Ha images were shifted and rotated to match the jet features in neighboring SOT images.
\stereo EUVI images were processed using the \texttt{mk\_secchi\_map} routine that takes into account
the pointing information of the spacecraft, with the \texttt{/rotate\_on} option turned on to
make the solar north the $y$-axis. We did not attempt to align images from EUVI and other instruments, since
\stereo has a different distance and view angle to the Sun compared with the Earth view.
Meanwhile, the two \stereo spacecraft were too close to allow for triangulation in 3D geometry 	
at the time of the event.

The translation and rotation corrections found above were applied to all images
from the corresponding instruments. The overall accuracy are $1\arcsec$--$2\arcsec$
and  $2\arcsec$--$3\arcsec$ in the $x$ and $y$ directions, respectively.

\section{B.~\hsi Spectral Analysis}
\label{Append_sect_hsi-spec}

\begin{figure}[thbp]      
 \epsscale{0.4}	
 \plotone{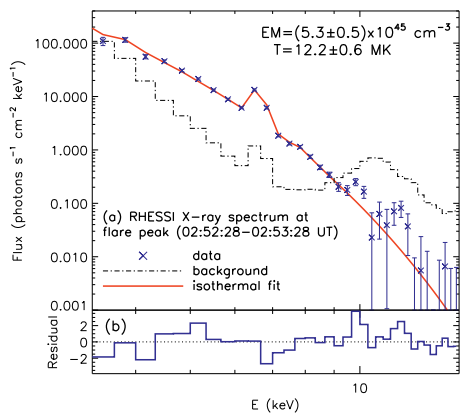}
 \caption[]{
 (a) Spatially integrated \hsi X-ray spectrum at the peak of the flare, averaged among detectors 1, 4, and 6 and fitted with an isothermal model.
 (b) Fitting residuals normalized to the 1$\sigma$ uncertainty of the measured flux at each energy.
 } \label{hsi_spec.eps}
 \end{figure}

To determine the temperature of the flare plasma, we fitted \hsi X-ray spectra.
Following \citet{KruckerS.Hinode-HSI.2007ApJ...671L.193K}, we used detectors 1, 4, and 6,
which had suffered the least	
radiation damage by the time of this event (5 years after the launch)
and had the best spectral resolution among the nine germanium detectors. We fitted the spectral data from each individual
detector separately, averaged the fitting parameters among the three detectors to give the final result,
and used their standard deviations as the uncertainties. Details of this procedure were described in
\citet{LiuW_2LT.2008ApJ...676..704L} and \citet{MilliganDennis.2009ApJ.EIS-v-T}.
Figure~\ref{hsi_spec.eps} shows the spectrum at the peak of the flare, which is best fitted with an isothermal model
and  yields a temperature of $T=12.2 \pm 0.6$~MK and emission measure of ${\rm EM}=(5.3 \pm 0.5) \E{45} \pcmc$.
There is pronounced ion line emission at 6.7~keV and no signature of nonthermal emission.
%

\section{C.~Nature of \CaIIH Emission Displacements}
\label{Append_sect_Ca-displace}

The resonance H line emission of \ion{Ca}{2} ions has contributions from scattering
of photospheric radiation and from thermal emission (collisional excitation).
In general, the former decreases with temperature because of progressive ionization 
and the latter increases with temperature because of increased collisional rates.
The emission decreases sharply with temperature when $T \gtrsim 2\E{4} \K$
\citep{Gouttebroze.prominence-CaII-lines.1997SoPh..172..125G}.

Based on the following reasonings, we believe that the displacements of Ca emission features presented in
this paper primarily result from mass motions rather 
than sequential excitations of emission due to temperature or density variations.
The \CaIIH line emitting plasma has typical temperatures of $\lesssim$$2\E{4} \K$,
which is in the temperature range of the chromosphere.
The Ca material bundle and loop also first appear at spicule heights in the chromosphere and then develop upward.
This suggests a chromospheric origin of the material that is likely of high density
as a result of flux emergence discussed in Section~\ref{sect_discuss}.
In addition, such features are no weaker in \CaIIH emission and in 195~\AA\ absorption
than the nearby prominence (Figure~\ref{sot_trace_xrt.eps}) at similar temperatures. 
This prominence, clearly visible in AR~10940 through its entire disk passage
({\footnotesize \href{http://www.solarmonitor.org/index.php?date=20070206}{http://www.solarmonitor.org/index.php?date=20070206}}),
is now just behind the limb and in the background of the event. Assuming that they have
comparable line-of-sight extents, the density of the prominence, which has typical values of $10^{11}$--$10^{12}\pcmc$,
provides a lower limit for the density in the Ca material bundle, loop, and jet.
This inference has two implications:
 (1) the temperature of the dense, cool material distributed in such an extended volume 
(cf., the compact flare site) is not expected to change rapidly because of its large heat capacity,
and thus the motions of these Ca emission features are unlikely a result of sequential temperature variations
on such short time scales of $\sim$10~s; 
and (2) more importantly, these features are two orders of magnitude denser than the ambient corona,
and thus their motions are unlikely a consequence of sequential compression of coronal plasma either. 
Moreover, compression would heat the million degree coronal plasma to even higher temperatures 
that are further away from the \CaIIH emission regime of two orders of magnitude cooler. 
For comparison, in an MHD experiment \citep{PariatE.twist-jet-MHD.2009ApJ...691...61P}, 
the compressional density enhancement responsible for supplying mass to a jet from a coronal origin
is merely a factor of two. 


\end{document}